\documentclass[letterpaper]{article} 
\usepackage{aaai2026}  
\usepackage{times}  
\usepackage{helvet}  
\usepackage{courier}  
\usepackage[hyphens]{url}  
\usepackage{graphicx} 
\usepackage{graphicx}
\usepackage{natbib}  
\usepackage{caption} 
\usepackage{algorithm}
\usepackage{algorithmic}

\usepackage{newfloat}
\usepackage{listings}
\DeclareCaptionStyle{ruled}{labelfont=normalfont,labelsep=colon,strut=off} 
\floatstyle{ruled}
\newfloat{listing}{tb}{lst}{}
\floatname{listing}{Listing}
\usepackage{graphicx}
\usepackage{url}
\usepackage{booktabs}
\usepackage{caption}
\usepackage{subcaption}

\newtheorem{theorem}{Theorem}
\usepackage{amsmath}
\usepackage{amssymb}
\usepackage{amsfonts}

\title{Physical‑regularized Hierarchical Generative Model for Metallic Glass Structural Generation and Energy Prediction}
\author {
    Qiyuan Chen\textsuperscript{\rm 1},
    Ajay Annamareddy\textsuperscript{\rm 1},
    Ying-Fei Li\textsuperscript{\rm 2},
    Dane Morgan\textsuperscript{\rm 1}, 
    Bu Wang\textsuperscript{\rm 1}\thanks{corresponding author}
}
\affiliations {
    \textsuperscript{\rm 1}University of Wisconsin - Madison\\
    \textsuperscript{\rm 2}Stanford University\\
    bu.wang@wisc.edu
}

\usepackage{bibentry}

\begin{document}

\maketitle

\begin{abstract}

Disordered materials such as glasses, unlike crystals, lack long‑range atomic order and have no periodic unit cells, yielding a high‑dimensional configuration space with widely varying properties. The complexity not only increases computational costs for atomistic simulations but also makes it difficult for generative AI models to deliver accurate property predictions and realistic structure generation. In this work, we introduce \textsc{GlassVAE}, a hierarchical graph variational autoencoder that uses graph representations to learn compact, translation‑, and permutation‑invariant embeddings of atomic configurations.  The resulting structured latent space not only enables efficient generation of novel, physically plausible structures but also supports exploration of the glass energy landscape. To enforce structural realism and physical fidelity, we augment \textsc{GlassVAE} with two physics‑informed regularizers: a  radial distribution function (RDF) loss that captures characteristic short‑ and medium‑range ordering and an energy regression loss that reflects the broad configurational energetics. Both theoretical analysis and experimental results highlight the critical impact of these regularizers. By encoding high‑dimensional atomistic data into a compact latent vector and decoding it into structures with accurate energy predictions, \textsc{GlassVAE} provides a fast, physics‑aware path for modeling and designing disordered materials.
\end{abstract}

\begin{links}
    \link{Code \& Data}{https://github.com/EricCH97/GlassVAE}
\end{links}

\section{Introduction}

Atomistic simulations of disordered materials—glasses in particular—are computationally intensive. The absence of long-range periodicity in these systems prevents the use of compact unit-cell descriptions, making new configurations have to be generated from first principles.\cite{bamerMolecularMechanicsDisordered2023, jungRoadmapMachineLearning2025} While generative AI has been successfully applied to crystalline solids \cite{court3DInorganicCrystal2020, liGenerativeDesignCrystal2025, longConstrainedCrystalsDeep2021, nouiraCrystalGANLearningDiscover2019} and molecules by using structure representations like the Crystallographic Information File (CIF) \cite{flam-shepherdLanguageModelsCan2023, zeniMatterGenGenerativeModel2024} and Simplified Molecular Input Line Entry System (SMILES)\cite{oboyle_deepsmiles_2018, skinnider_invalid_2024, weininger_smiles_1988, sun_ice_2024}, yet, no equivalent representation exists for inorganic glasses. Consequently, researchers rely on local geometric fingerprints—such as smooth overlap of atomic positions (SOAP) kernels \cite{de_comparing_2016, lin_expanding_2024, bartok_representing_2013}, or atom-centered symmetry functions (ACSF) \cite{mudassir_systematic_2022} —which capture only short‑range order and omit system‑level geometry and related properties. Generative models trained solely on these local features often propose structures that appear locally plausible yet not satisfying global property constraints, hindering the discovery of new structures and materials.

In addition, glasses are inherently non-equilibrium systems with complex, high-dimensional configurational energy landscapes\cite{debenedetti_supercooled_2001}. Exploring these landscapes to identify realistic glassy structures is computationally intensive, as such structures occupy only a tiny subset of all possible configurations. Accurate generation of glass structures therefore requires the modeling to be aware of their configurational energy. Traditionally, molecular dynamics (MD) simulations have been the primary tool, where atomic interactions are explicitly simulated. By replicating the melt-quench process using MD, specific regions of the energy landscape are explored and yields structures resembling experimentally produced glasses. However, due to computational constraints, MD quenching rates are typically 6–9 orders of magnitude faster than the experimental rates\cite{binder_molecular_2004}. This results in structures with lower thermodynamic stability and altered properties\cite{yu_structural_2021}. This timescale mismatch is also a major challenges for simulating key glass behaviors such as relaxation and aging\cite{yu_stretched_2015, berthier_modern_2023}.

To address these challenges, graph-based representations combined with generative frameworks capable of predicting configurational energy offer a promising approach for modeling disordered systems like glasses. In these representations, atoms are treated as nodes and interatomic distances serve as invariant edge features. Through symmetric aggregation, such models inherently respect E(3) symmetries—translation, rotation, and permutation invariance—enabling them to encode essential physical constraints directly into the architecture.\cite{satorras_en_2022, bao_equivariant_2023} This built-in invariance allows for the compression of high-dimensional atomic configurations into a compact, continuous latent space, significantly reducing computational overhead while retaining critical structural information. An explicit latent space is particularly useful for studying glass systems. Traversing this latent space enables the generation of new glassy structures and systematic exploration of their thermodynamic properties.

Despite the strong predictive performance of graph neural networks (GNNs) for atomic-scale property estimation \cite{liPredictingInterpretingEnergy2024}, embedding them within a generative modeling framework remains nontrivial. Hierarchical latent-space variational autoencoders (H-VAEs) \cite{klushynLearningHierarchicalPriors2019, vahdatNVAEDeepHierarchical2020, yingHierarchicalGraphRepresentation2019} overcome this by learning multiscale latent representations: high‑level variables capture global composition and medium to long-range features, while low‑level variables encode local geometry. This architecture enhances expressiveness, mitigates posterior collapse, and produces physically plausible atomistic structures.

Here, we introduce \textsc{GlassVAE}, a novel hierarchical variational autoencoder specifically designed to compress high-dimensional atomic graphs of disordered systems into a structured, low-dimensional latent space. \textsc{GlassVAE}'s effectiveness stems from two key innovations: (i) Physics-Informed Regularization:  A radial-distribution-function (RDF) loss ensures the preservation of essential short- and medium-range structural order \cite{rapaport_art_2004, allen2017computer}, while an energy-regression loss aligns the latent embeddings with thermodynamic reality. (ii) Specialized Hierarchical Latent Structure: A graph-level latent variable captures global characteristics such as composition and the overall energy landscape, while a distinct edge-level latent variable refines local geometric details. This strategic division of responsibilities within the hierarchy effectively balances global energetic considerations with local structural fidelity, without unduly inflating the dimensionality of the latent space.

In this work, we demonstrate a prototype \textsc{GlassVAE} on the CuZr metallic glass system. This system is selected to illustrate both the potential and the complexity of disordered alloys.\cite{pauly_transformation-mediated_2010} These amorphous metals combine the high strength of crystalline metals with the formability of polymers. \cite{schroers_thermoplastic_2011} Their non‑crystalline atomic arrangement underlies exceptional mechanical, thermal, and corrosion‑resistance properties, driving their application in diverse fields including aerospace components, biomedical implants, and advanced sports equipment.\cite{trexler_mechanical_2010, bansal_chapter_1986} Despite extensive study, the details of their structure-property relationship are still not well understood. \cite{starr_what_2002, cao_structural_2018} Through theoretical analysis and experiments on CuZr metallic glass, we demonstrate that \textsc{GlassVAE} accurately reproduces potential-energy distributions, achieves low root-mean-square error (RMSE) in reconstructing atomic positions, and, crucially, generates novel configurations that simultaneously adhere to RDF profiles and energetic constraints. By effectively unifying geometric and energetic priors within a single, tractable latent space, \textsc{GlassVAE} establishes the first end-to-end generative framework specifically tailored for disordered inorganic systems. This work opens a new avenue for AI-guided discovery and design of complex glassy materials.

\section{Related Work}
\textbf{Generative AI in Materials Science} \ GenAI models—including Generative Adversarial Networks (GANs) \cite{nouira_crystalgan_2019, xuGenerativeAdversarialNetworks2023}, Variational Autoencoders (VAEs)\cite{gomez-bombarelli_automatic_2018, luo_deep_2024}, autoregressive models like Transformers \cite{wang_compositionally_2021}, and diffusion models \cite{satorras_en_2022, bao_equivariant_2023}—have become powerful tools in materials design. These approaches are increasingly applied to sample novel compositions, predict structures, and estimate properties, often within a unified computational framework \cite{schleder_dft_2019}. Significant breakthroughs have been demonstrated, particularly for crystalline materials, \cite{xie_graph_2019, zeniMatterGenGenerativeModel2024, ren_invertible_2022} and for organic molecules, enabling de novo drug design and property optimization\cite{ gomez-bombarelli_automatic_2018, hoogeboom_equivariant_2022}. Despite these successes, most generative efforts remain confined to systems exhibiting well-defined order, such as the periodic arrangements in crystals or the discrete bonding patterns in molecules. While generative techniques have been applied to glasses, \cite{li_inverse_2024, cassar_designing_2021}these studies have largely focused on predicting macroscopic properties, generating microstructure morphologies, rather than atomic-level configurations \cite{jungRoadmapMachineLearning2025}.

\textbf{Graph Neural Networks} \ Graph neural networks (GNNs) have become indispensable tools in molecular and materials simulation, enabling accurate prediction of energies, forces, and other properties \cite{gilmer_neural_2017, satorras_en_2022, sanchez-gonzalez_learning_2020}.  Among all, Wang et al.,\ combined GNNs with swap Monte Carlo to inversely design Cu–Zr metallic glasses with enhanced plastic resistance \cite{wang_inverse_2021}, and Li et al.,\ applied GNNs to predict energy barriers and interpret the behavior of metallic glasses \cite{liPredictingInterpretingEnergy2024}.  Yoshikawa et al.,\ integrated self‑attention into a GNN for structural classification of glass‑forming liquids \cite{yoshikawa_graph_2025}, while Bapst et al.,\ trained a GNN solely on initial particle positions to predict long‑time glass dynamics \cite{bapst_unveiling_2020}. Although these studies demonstrate the power of GNNs for property prediction and classification in glassy systems, none simultaneously address structural generation.  Our approach fills this gap by learning an internal generative mechanism that bridges state‑of‑the‑art molecular‑graph generative models with physically consistent atomic reconstructions.

\textbf{Physical-informed Descriptors} \ Global, physics‑informed descriptors—such as Crystallographic Information Files (CIFs) for crystals \cite{flam-shepherdLanguageModelsCan2023, zeniMatterGenGenerativeModel2024} and SMILES for organic molecules \cite{weininger_smiles_1988, oboyle_deepsmiles_2018, skinnider_invalid_2024}—require long‑range order and thus fail for glasses. Instead, local geometric fingerprints like smooth overlap of atomic positions (SOAP) kernels \cite{bartok_representing_2013, de_comparing_2016, lin_expanding_2024} or atom‑centered symmetry functions (ACSF) \cite{mudassir_systematic_2022} capture only short‑range structure and omit system‑level properties such as total potential energy. The radial distribution function (RDF) encodes short‑ and medium‑range order in disordered materials.  For example, Krykunov et.al.,\ introduced atomic property–weighted RDFs with analytic Gaussian kernels that both outperform and accelerate Bag‑of‑Bonds for atomization energy prediction \cite{krykunov_bond_2018}, and Watanabe et.al.,\ developed an on‑the‑fly RDF‑based sampling method to flag anomalous configurations and build robust water potentials \cite{watanabe_machine_2024}.  To our knowledge, we are among the first to integrate an RDF‑based loss into a variational graph model for glasses, combining RDF‑driven regularization with graph representations to capture both local geometry and global energetics.

\section{Physical-regularized Hierarchical Graph Variational Autoencoder}

In this work, we introduce \textsc{GlassVAE}, a generative framework for disordered metallic–glass configurations that jointly captures local atomic geometry and global energetics. Our approach addresses three core tasks:
(i) Reconstruction: Recover atom types, interatomic displacements and distances, and absolute positions;
(ii) Energy prediction: Estimate the total potential energy of each configuration;
(iii) Generation: Sample new, physically realistic atomic structures from the learned latent distribution. Formally, each molecular‐dynamics snapshot is represented by \((\mathbf{R},\mathbf{t},E)\), where
\(
  \mathbf{R} = \{\mathbf{r}_i\in\mathbb{R}^3\}_{i=1}^N
  \quad\text{(positions in a periodic box }\mathbf{L}=[L_x,L_y,L_z]\text{),}\)
\(\mathbf{t}\) is the one‐hot encoding of atom types, and \(E\) is the total potential energy. We learn
\(
  \mathcal{E} : (\mathbf{R},\mathbf{t},E) \;\mapsto\; \boldsymbol{z}
  ,\quad
  \mathcal{D} : \boldsymbol{z} \;\mapsto\; (\mathbf{R},\mathbf{t}),
  \quad \text{and}\quad 
  \mathcal{P} :  \boldsymbol{z} \;\mapsto\; \mathbf{E}
\)
so that the latent variable \(\boldsymbol{z}\) retains all essential physical information for faithful reconstruction, accurate energy estimation, and the generation of novel, physically plausible structures.

\subsection{Graph Representation and Invariance}
\label{graph_rep}

Physical interactions in materials depend only on atomic species and their relative arrangements, not on ordering, absolute position, or orientation. Therefore, atomistic representations and predictions must be invariant under permutation, translation, and rotation—swapping identical atoms, shifting all positions by \(\Delta t\), or applying any orthogonal rotation \(R\) leaves interatomic distances (and thus all physical properties) unchanged. To enforce these invariance, we convert each point cloud into an attributed graph \(\mathcal{G}=(V,E)\), where
Nodes \(i\in V\) carry one‑hot features \(t_i\) encoding atomic species i.e., \(x_i \in \{0, 1\}\); and 
Edges \((i,j)\in E\) connect atoms within a cutoff distance (under periodic boundaries) and carry attributes
  \[
    \mathbf{a}_{ij}=\bigl(\Delta\mathbf{r}_{ij},\,d_{ij}\bigr),\quad
    \text{where} \ \Delta\mathbf{r}_{ij}=\mathbf{r}_i-\mathbf{r}_j,\;
    d_{ij}=\|\Delta\mathbf{r}_{ij}\|_2.
  \]

A graph neural network with symmetric (e.g., mean) message‑passing layers aggregates these features, guaranteeing invariance to node ordering and rigid‑body transformations. We further support these invariance claims with a formal theorem.

\begin{theorem}[Invariance of the Graph Representation]
\label{thm:invariance}
Let an atomic configuration be given by positions $\mathcal{R} = \{\mathbf{r}_i\}_{i=1}^N \subset \mathbb{R}^3$ and one‐hot node features $\{\mathbf{t}_i\}$. Construct a graph $G=(V,E)$ by connecting all pairs with $\|\mathbf{r}_i - \mathbf{r}_j\|_2 \le \mathrm{CUTOFF}$, and assign each edge the attribute
\[
\mathbf{a}_{ij}
=\bigl(\Delta \boldsymbol{r_{ij}},\|\mathbf{r}_i - \mathbf{r}_j\|_2\bigr),
\quad
\text{where}\ \Delta \mathbf{r}_{ij} = \mathbf{r}_i - \mathbf{r}_j.
\]
Suppose the graph neural network updates and pools features using a symmetric (permutation‐invariant) aggregation function $S$ (e.g.\ mean or sum).  Then the resulting representation is invariant to\\
\textbf{Permutation of nodes:}  Since $S$ does not depend on the ordering of its inputs, any re-indexing of the atoms leaves the pooled feature unchanged.\\
\textbf{Translation of the entire structure:}  If $\mathbf{r}_i' = \mathbf{r}_i + \mathbf{t}$ for all $i$, then
  \[
    \Delta \mathbf{r}_{ij}' = (\mathbf{r}_i + \mathbf{t}) - (\mathbf{r}_j + \mathbf{t})
    = \mathbf{r}_i - \mathbf{r}_j
    = \Delta \mathbf{r}_{ij},
  \]
  so all edge‐wise inputs to the network are identical.\\
\textbf{Rotation of the entire structure:}  If $\mathbf{r}_i' = R\,\mathbf{r}_i$ for some orthogonal $R\in\mathbb{R}^{3\times3}$, then
  \[
    \Delta \mathbf{r}_{ij}' = R(\mathbf{r}_i - \mathbf{r}_j),
    \quad
    \|\Delta \mathbf{r}_{ij}'\|_2
    = \|R(\mathbf{r}_i - \mathbf{r}_j)\|_2
    = \|\Delta \mathbf{r}_{ij}\|_2,
  \]
  so all distance‐based edge features are preserved.
\end{theorem}

\noindent Theorem~\ref{thm:invariance} thus demonstrates that our graph construction method and the use of symmetric aggregation functions yield a representation of 3D atomic configurations that is inherently invariant to permutations, translations, and rotations—properties that are essential for reliable modeling of physical systems.

\subsection{Model Architecture}
\begin{figure*}[!htb]
    \centering    \includegraphics[width=0.7\linewidth]{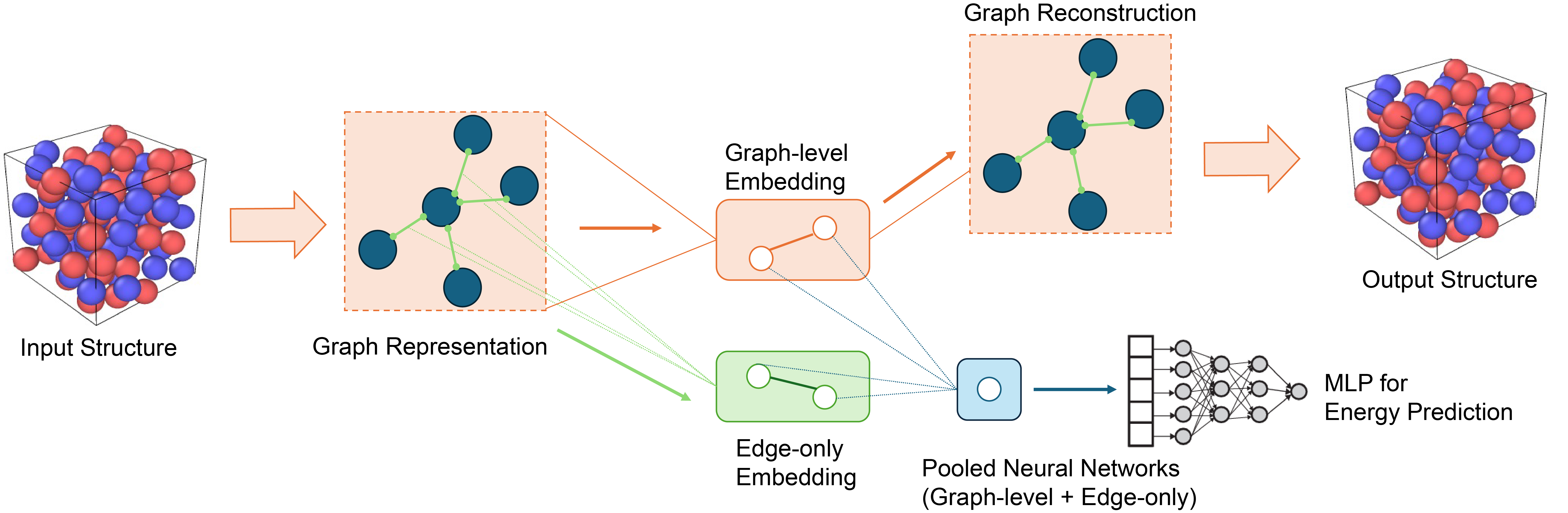}
    \caption{Schematic of Hierarchical Variational Autoencoder}
    \label{fig:1}
\end{figure*}

To jointly capture the intricate local geometry and global properties (e.g., energies) of disordered metallic glass configurations, we propose a hierarchical graph variational autoencoder (\textsc{GlassVAE}). Figure~\ref{fig:1} sketches its architecture. Sec. \ref{graph_rep} states how we build a graph starting from an atomic configuration (also see Sec. \ref{Algo}, \ref{sec:training} for more details). From graph representations, we have 

\textbf{Dual-path Encoder} \ The dual-path encoder that distills two complementary summaries.  The graph-level encoding applies \(L\) layers of message passing to embed local neighbourhoods yielding a graph vector that two linear heads convert into the variational parameters \(\mu\) and \(\log\sigma^{2}\). We could then reparameterize them into a decoder head \(z \sim \mu + \sigma\). In parallel, the edge path feeds all \(a_{ij}\) through a residual stack of \(L_e\) blocks and averages the refined edge features into an edge descriptor \(\boldsymbol{s}\). 

\textbf{Decoder} \ The decoder maps $\mathbf z$ back to atomic species,i.e., \( \hat{x} = D_1(z)\) and edge attributions, i.e, \(\hat{a_{ij}} = D_2(z)\) thereby reconstructing the full configuration.

\textbf{Property Predictor} \ A separate fusion head consumes the concatenated vector \(\Phi(\boldsymbol{z} \!\parallel\! \boldsymbol{s})\) to predict the total energy $\hat E$, allowing the model to learn structure–property relationships directly from the latent space.\\

\textbf{Loss Function} \ The overall training objective is a weighted sum of several loss components:
  
\begin{equation}
\begin{aligned}
\mathcal{L} = \; \alpha_{\text{node}} \, \mathcal{L}_{\text{node}} + \alpha_{\text{edge}} \, \mathcal{L}_{\text{edge}} + \alpha_{\text{energy}} \, \\ \mathcal{L}_{\text{energy}} + \beta_{\text{KL}} \, D_{\text{KL}} + \alpha_{\text{RDF}} \, \mathcal{L}_{\text{RDF}}.
\end{aligned}
\label{eq:loss function}
\end{equation}

where each term enforces a different aspect of structural fidelity:

\(\mathcal{L}_{\mathrm{node}}\) is cross‑entropy between ground‑truth one‑hot atom types and predicted class probabilities;
\(\mathcal{L}_{\mathrm{edge}}\) is MSE on predicted distances plus a cosine‑similarity penalty to preserve directional consistency of edge features;
\(\mathcal{L}_{\mathrm{energy}}\) is MSE between predicted and true potential energies;
\(D_{\mathrm{KL}}\) is Kullback-Leibler (KL) divergence;
\(\mathcal{L}_{\mathrm{RDF}}\) is \(\ell_2\) distance between soft histograms of predicted and true pairwise distance distributions.Details about how we calculated the loss can be referred to Sec.\ref{sec:training}.

\subsection{Generation}
\label{generation_discuss}

Generation is one of the key features of \textsc{GlassVAE}: its structured latent space enables both random and energy‐conditioned synthesis of new, physically realistic configurations through the decoder. Here, we outline the two modes of operation. We also list the results of generation in Sec. \ref{sampling from latent space}.

\textbf{Random generation.} To produce novel structures, we draw
\[
  \boldsymbol{\epsilon} \sim \mathcal{N}(\mathbf{0}, \mathbf{I}),
\text{and set} \
  \mathbf{z} = \boldsymbol{\mu}(x) + \boldsymbol{\sigma}(x)\,\odot\,\gamma\boldsymbol{\epsilon},
\]
where “\(\odot\)” denotes element-wise multiplication. This turns sampling
into a differentiable operation with respect to \(\boldsymbol{\mu}\) and \(\boldsymbol{\sigma}\).
here \(\boldsymbol{\epsilon}\) is a small perturbation. Then the sampled new structures could be decoded through \(\hat{x}=D_1(z) \ \text{and} \ \hat{a} = D_2(z)\).

\medskip
\textbf{Conditional generation.}
To constrain the generated energy to \([E_{\min},E_{\max}]\), we could use conditional generation. Formally, we assume a standard normal prior \(p(z) \sim \mathcal{N}(0,I)\). To enforce \(\hat E(z)\in [E_{\min},E_{\max}]\), we initialize \(z\) at the encoder’s posterior mean and minimize
\begin{align*}
    \mathcal{L}(z)
= & \bigl[\max(E_{\min}-\hat E(z),0)\bigr]^{2} 
 + \\ & \bigl[\max(\hat E(z)-E_{\max},0)\bigr]^{2}
+\lambda\|z\|_{2}^{2}
\end{align*}

for \(T\) steps. Decoding the optimized code \(z^{*}\) yields a structure whose energy lies within the target range.

\subsection{Theoretical Contribution of Physics-Informed Regularizer}
\label{subsec:phys_reg_theory}

Physics-informed loss terms (i.e.\ energy and RDF in our case, Eq.\ref{eq:loss function}) do more than improve reconstruction accuracy—they can associate the latent variables to physically admissible manifolds.  Below we formalize a theoretical explanation: imposing an operator‐based regularizer \(\|\mathcal{A}\boldsymbol{u}-v\|_{L^{2}}\) induces a bound on the latent test error that decays with the training error and the quadrature resolution.  In other words, every unit of training progress achieved by a physics-informed term translates directly into a proportionate tightening of the latent-space generalization gap (see full proof in Sec. \ref{the2_full}).

\vspace{0.5em}
\noindent\textbf{Functional setting.}
Let \(\mathcal{U}=W^{s,q}(\mathbb{R}^{3})\) and \(\mathcal{V}=L^{p}(\mathbb{R}^{3})\) with \(1\!\le\!p,q\!\le\!\infty\) and \(s\!\ge\!0\).
Given a neural approximation \(\boldsymbol{u}(\cdot;\theta)\!\in\!\mathcal{U}\), the operator
\(\mathcal{A}:\mathcal{U}\!\to\!\mathcal{V}\) extracts a physical observable  
(e.g.\ predicted RDF, energy density), while \(v\!\in\!\mathcal{V}\) encodes the target dictated by first-principles or simulation data.

\noindent\textbf{Assumption 1 - Operator stability}
\label{assump:op-stab}

There exists \(C_1>0\) such that
\(\|\boldsymbol{u}-\boldsymbol{u}'\|_{\mathcal{U}}
      \le C_1\|\mathcal{A}\boldsymbol{u}-\mathcal{A}\boldsymbol{u}'\|_{\mathcal{V}}\)
for all \(\boldsymbol{u},\boldsymbol{u}'\in\mathcal{U}\).

\noindent\textbf{Assumption 2 - Quadrature accuracy}\label{assump:quad}

A numerical quadrature with nodes \(\{\mathbf{x}_i,w_i\}_{i=1}^{N}\) satisfies  
\(|\int g-\sum_{i}w_ig(\mathbf{x}_i)|\le C_2 N^{-\alpha}\)
for all \(g\in\mathcal{V}\), some \(C_2>0\) and \(\alpha>0\).

\vspace{0.25em}
\noindent\textbf{Train–test decomposition.}
Define the physics-informed training error
\[
\mathcal{E}_{\mathrm{train}}(\theta)
      :=\sum_{i=1}^N w_i\bigl(\mathcal{A}\boldsymbol{u}(\mathbf{x}_i;\theta)-v(\mathbf{x}_i)\bigr)^2,
\]
and let \(\theta^\star\) be a minimizer.  The latent test error is
\(\mathcal{E}_{\mathrm{test}}
      :=\|\boldsymbol{u}(\cdot;\theta)-\boldsymbol{u}(\cdot;\theta^\star)\|_{\mathcal{U}}\).

\begin{theorem}[Generalization bound]\label{thm:gen}
Under Assumptions \ref{assump:op-stab}–\ref{assump:quad}, we have
\[
\mathcal{E}_{\mathrm{test}}
      \;\le\;
      C_1\mathcal{E}_{\mathrm{train}}^{1/2}
      +C_1\,C_2^{1/2}N^{-\alpha/2}.
\]
\end{theorem}

\begin{table*}
\centering
\resizebox{0.9\linewidth}{!}{
\begin{tabular}{@{}lccccc@{}}
\toprule
\textbf{Method} & \multicolumn{2}{c}{\textbf{Energy (eV/atom)}} & \multicolumn{2}{c}{\textbf{Paired Distances(\r{A})}} & \textbf{Node} \\
 &  RMSE &  $R^2$ &  RMSE & $R^2$  & BCE\\
\midrule
\texttt{Distance VAE (distance matrix as input)} &  11.85 & 0.89 & 0.24 & 0.77 & --- \\
\texttt{GNN + MLP (no generation)} &  2.15 &  0.98 &  --- & --- & ---\\
\texttt{Single Graph-VAE (single latent space)} &  29.32 & $<$ 0 & 0.18 & 0.93 & $<$ 0.1 \\
\texttt{GraphVAE (no RDF)} &  0.37 & $>$ 0.99 & 0.14 & 0.94 & $<$ 0.1 \\
\textbf{GlassVAE (best, + RDF)} &  \textbf{0.32} & \textbf{ $>$ 0.99} & \textbf{0.025} & \textbf{$>$ 0.99} & \textbf{0.091}\\
\bottomrule
\end{tabular}
}
\caption{Performance on the test set. We report RMSE (lower is better), $R^2$ (higher is better), and Binary Cross Entropy (BCE) (lower the better).}
\label{tab:main_results}
\end{table*}

\begin{figure*}[!htb]
    \centering
\includegraphics[width=0.8\linewidth]{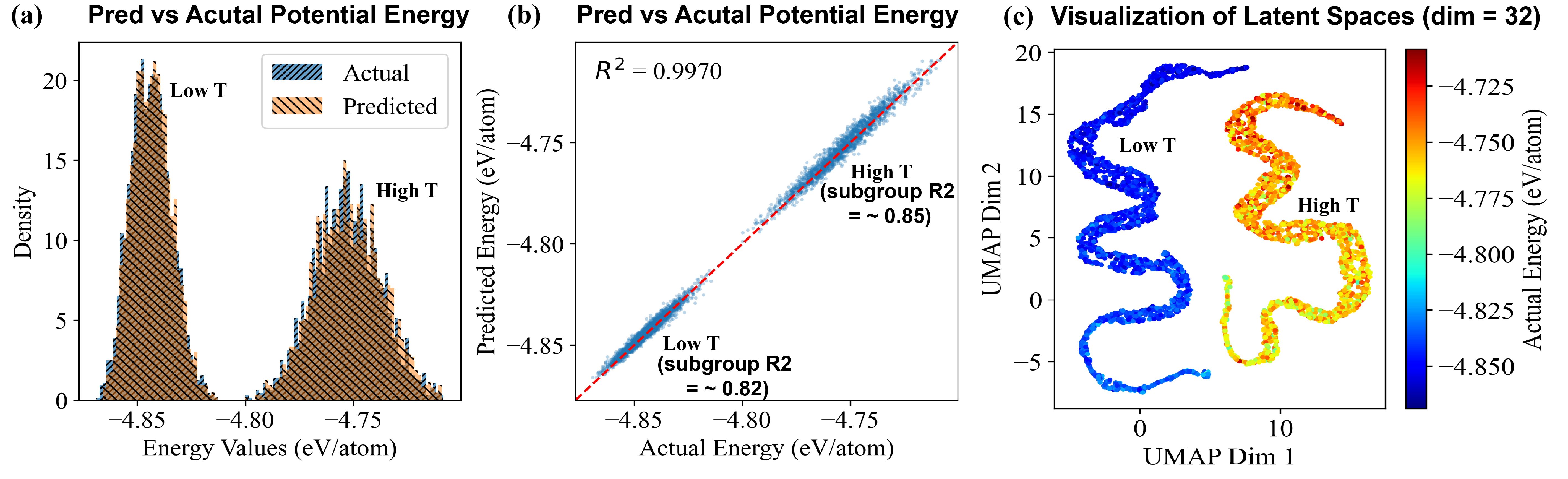}
    \caption{Energy Predictions and Latent Space Visualization. (a-b) Predicted vs actually potential energy; (c) visualization of graph latent spaces via UMAP projection}
    \label{fig:energy}
\end{figure*}

Because both our energy‐ and RDF‐based penalties satisfy the Lipschitz‐stability condition (Assumption~1) and are evaluated in low-dimensional space ($D$ = 3), each physics‐informed term directly drives down \(\mathcal{E}_{\mathrm{train}}\).  By minimizing these terms we therefore \emph{provably tighten} the upper bound on \(\mathcal{E}_{\mathrm{test}}\), improving generalization to unseen configurations. In this way, our physical regularizer are more than heuristics—they provide a principled mechanism for guaranteeing better predictive performance. This theoretical picture also has evidence from our experiments.  Figure~\ref{fig:energy}(c) visualizes the latent space of the full \textsc{GlassVAE}, whereas Fig.~\ref{fig:simpleVAE w/o Physical} shows the same model \emph{without} energy and RDF regularizes.  The baseline VAE collapses into a tight cluster that fails to separate high- and low-energy structures.  By contrast, the physics-aware \textsc{GlassVAE} learns a more expressive, well-spread latent landscape that clearly distinguishes configurations by energy—exactly the discriminative structure predicted by the theorem.

\section{Experiments and Results}

To evaluate the performance of \textsc{GlassVAE}, we trained the model on molecular‐dynamics (MD) trajectories of a Cu$_{50}$Zr$_{50}$ metallic glass and then assessed it on an unseen test set. The system comprised 108 atoms with a glass‐transition temperature $T_g = 700\,$K, with atomic interactions described using the embedded atom method (EAM) potential. The system was first equilibrated in the liquid state at 2000 K for 2 ns, then quenched to 1000 K in 50 K intervals at a cooling rate of 100 K per 6 ns and further cooled down in 20 K intervals at the rate of 100 K per 60 ns, all using the NPT ensemble. The simulations consisted of 108 atoms, and an MD time step of 1 fs is used. Periodic boundary conditions are employed in all three directions. We employed trajectories under 700 K, 760 K, 840 K, 920 K, 1000 K, with a total samples number of 50,000.  The in-detailed data pre-processing pipeline are introduced in Sec.\ref{dataprocess}. In order to show the scalability of performance of \textsc{GlassVAE}, we also evaluated \textsc{GlassVAE} on two independent Cu$_{50}$Zr$_{50}$ metallic‑glass trajectories at 720K and 960K, both initiated from distinct starting configurations (see details in Sec.\ref{dimension check and scalability} and \ref{fig:scalarable_energy}).

In summary, \textsc{GlassVAE}’s hierarchical latent space combined with physics‑informed regularization provides the flexibility needed for complex glassy systems and outperforms standard VAE variants (see Table.\ref{tab:main_results}), achieving superior accuracy in both property prediction and graph reconstruction. We show the detailed experimental results below and discuss them in more detail in the next section.

\subsection{Energy Prediction and Latent Space Exploration}

Another core aim of \textsc{GlassVAE} is accurate potential‐energy prediction. On the test set (Figs.~\ref{fig:energy}a–b), our model achieves an $R^2>0.99$, producing energy estimates whose overall distribution is virtually indistinguishable from the ground truth; even within defined subgroups (Fig.~\ref{fig:subgroup energyl}), it maintains $R^2\approx0.85$.  Overall, we achieved a $RMSE = 0.32$ among test dataset, which means on average a $\sim$ 0.025 eV reconstruction error on average. A UMAP projection of the latent codes (Fig.~\ref{fig:energy}c), colored by true energy, reveals two well‐separated clusters corresponding to configurations below and above the glass‐transition temperature \cite{mcinnes_umap_2020}. In contrast, the VAE without physics‐informed regularization (Fig.~\ref{fig:simpleVAE w/o Physical}) collapses into a single undifferentiated cluster. These results demonstrate that jointly training on energy not only delivers state‐of‐the‐art accuracy but also embeds meaningful energetic structure into the latent space, enabling clear separation of physical regimes in accordance with our theoretical findings (Sec.~\ref{subsec:phys_reg_theory}).

\subsection{Graph Reconstruction}

\begin{figure*}[!htb]
    \centering
\includegraphics[width=0.5\linewidth]{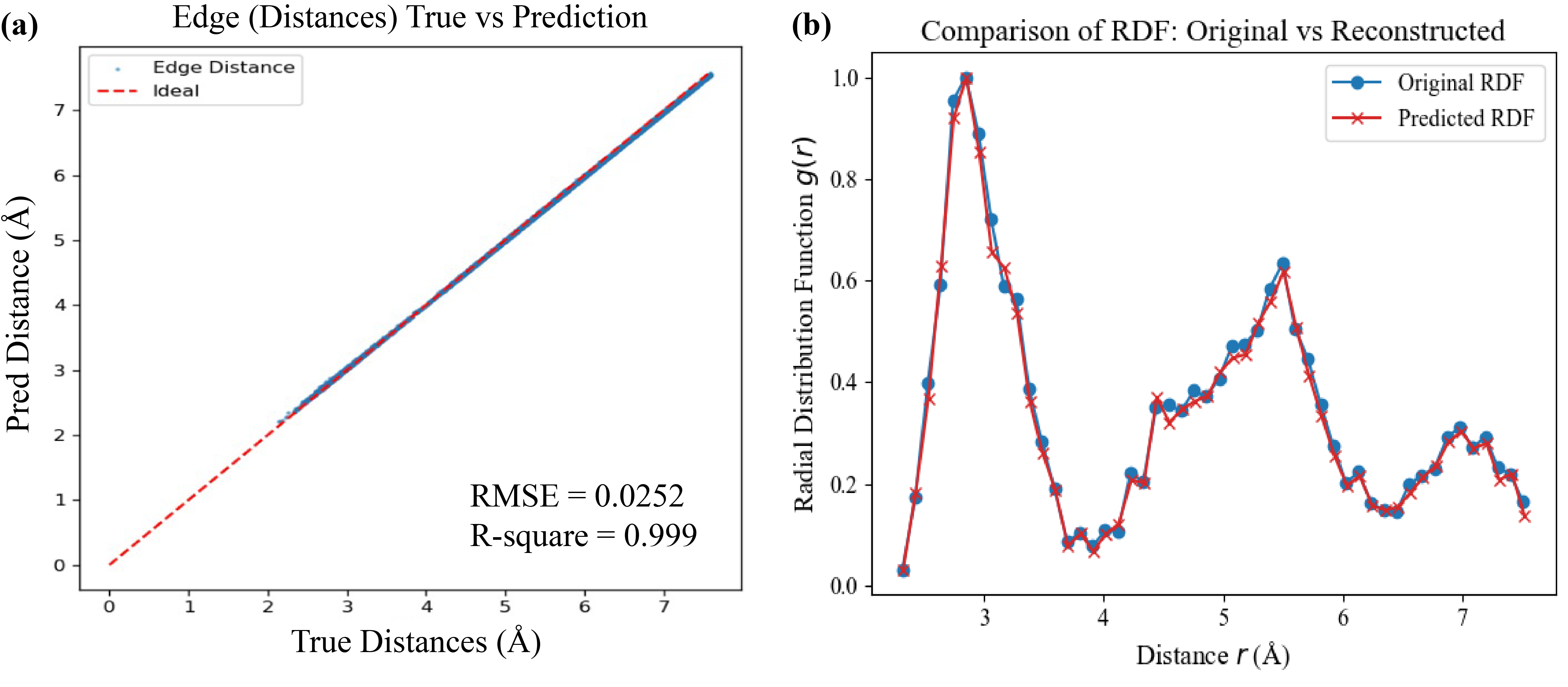}
    \caption{Edge Reconstruction Example (Distance unit \r{A}). (a) shows the reconstructed distributions v.s. the ground truth. (b) Comparison of original RDF v.s. reconstructed}
    \label{fig:edge}
\end{figure*}

In order to explore the reconstruction accuracy from \textsc{GlassVAE}, we primarily looked into the correct identification of atomic species and the accurate recovery of their pairwise \(L_2\) distances. Accordingly, we evaluate whether our model can faithfully reconstruct both node features (atomic types) and edge attributes (interatomic distances). Overall, we achieved a $RMSE = 0.025$ among test dataset, which means on average a $\sim$ 0.025 \r{A} reconstruction error on average. Fig. \ref{fig:edge} shows an example structure in the testing dataset. The model achieves excellent fidelity: it recovers atomic species with over 95 \% accuracy, it does a good job in predicting the distances with a over 0.99 $R^2$ and a rather small RMSE. Fig. \ref{fig:edge} (b) also exhibits the reconstructed distances follows a close similar distributions as original input. These findings confirm that \textsc{GlassVAE} preserves both the structural and physical information of the original graph throughout the decoding process.

\subsection{Sampling from the Latent Space}
\label{sampling from latent space}

\begin{figure*}
    \centering
\includegraphics[width=0.6\linewidth]{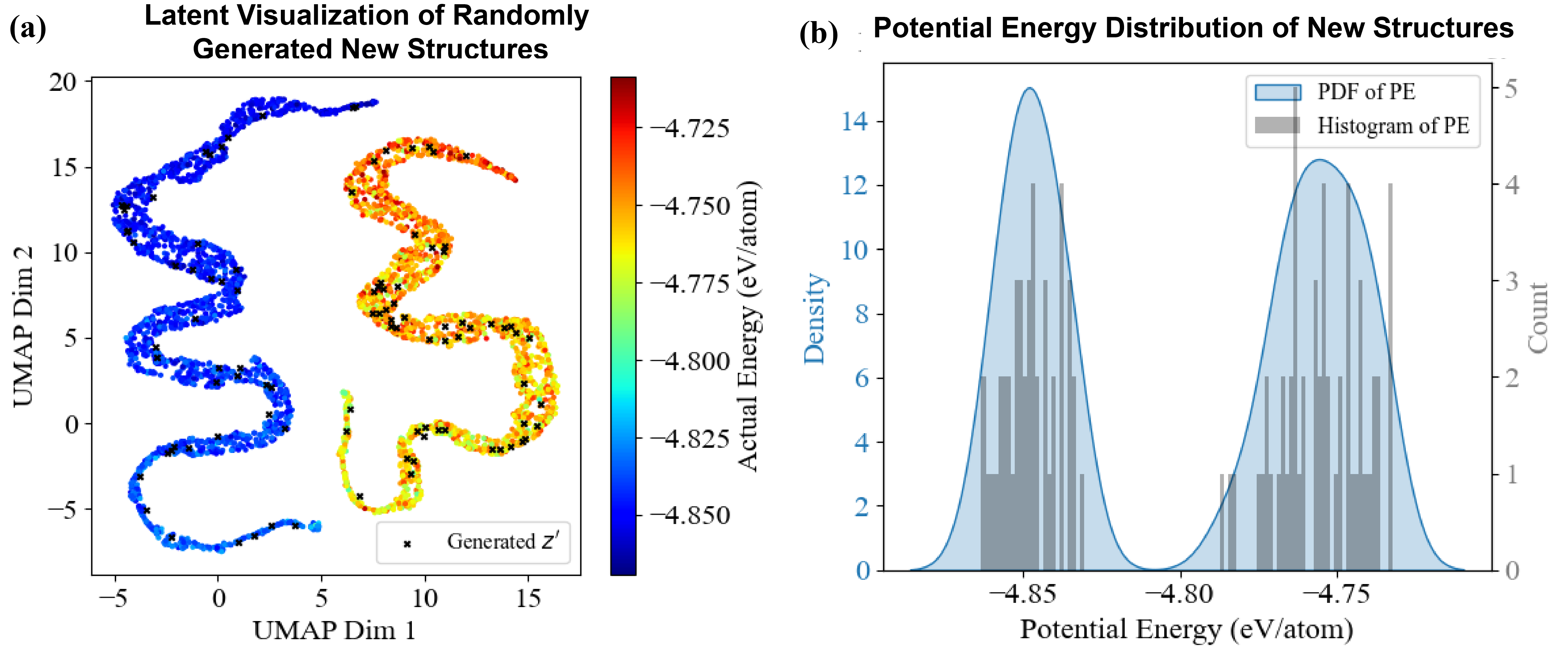}
    \caption{New Structure Generation. (a) Visualization of random sampled new structures in latent spaces (marked as black boxes); (b) Distribution of calculated potential energy of new structures, a kernel density estimate (KDE) was plot to show the continuous probability density (PDF) of new structures' potential energies}
    \label{fig:generation}
\end{figure*}

A key advantage of VAEs is their ability to generate novel samples. Here, we demonstrate these generative capability via both conditional and random sampling:
\textbf{Random Sampling} \ Random generation involves sampling from a simple standard normal distribution and then transforming this sample using learned mean and variance parameters, effectively drawing from the complex distribution learned by the model. Specifically, we sample directly from the standard normal distribution (the prior). We can also introduce controlled variations by adding a small amount of scaled noise to this initial sample before decoding. This perturbed sample is then passed through the decoder network to produce a final structure that represents a smooth deviation from the unperturbed version. Fig. \ref{fig:generation} (a - b) show the results of 100 randomly generated samples and their calculated potential energy distributions. Fig. \ref{fig:generation} (a) visualizes the new structures in latent space (cluster near known trajectories), indicating the generation of similar but novel structures. Moreover, the calculated potential energies of these generated structures (Fig. \ref{fig:generation} (b)) align well with the energy distribution of the training data, suggesting they are physically meaningful.

\begin{figure}
  \centering
  \includegraphics[width=0.6\linewidth]{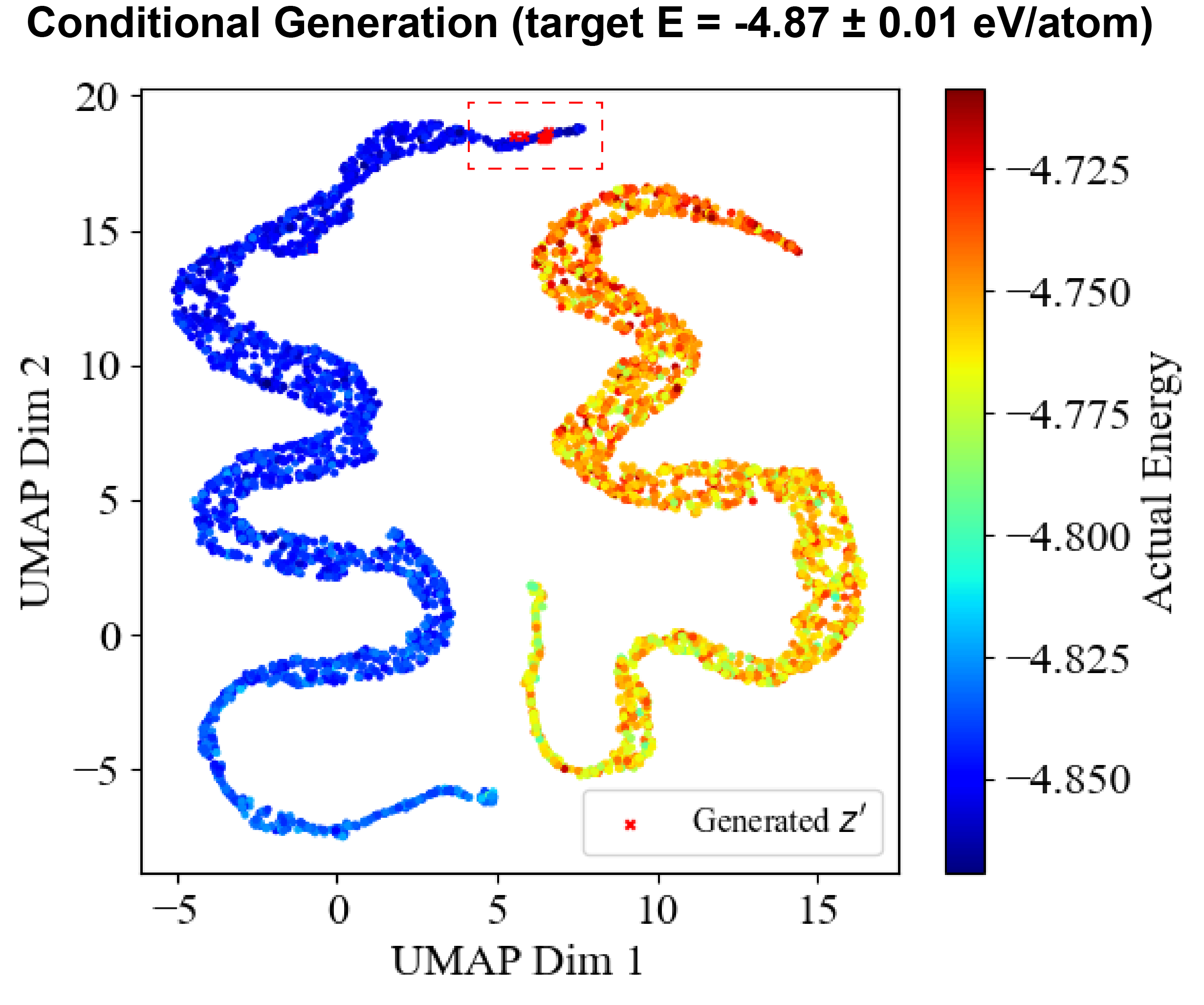}
  \caption{Conditional generation at $E_{\rm tar}=-4.87\pm0.01\ \mathrm{eV/atom}$. New structures are marked as red cross inside the labeled box}
  \label{fig:conditional_gen}
\end{figure}

\textbf{Conditional Generation} \ Instead, we could also generate atomic configurations whose potential energies lie within a specified interval by performing gradient‐based refinement directly in the \textsc{GlassVAE} latent space. Figure~\ref{fig:conditional_gen} illustrates this process for a target energy of \(E = -4.87 \pm 0.01\ \mathrm{eV/atom}\). The five generated structures are marked in latent space as red \('\times'\) symbols.
Both conditional generation and latent space optimization leverage the co-training of the energy prediction head to structure the latent space. By enforcing an energy constraint, as we discussed earlier in Sec. \ref{generation_discuss}, the generated configurations are more likely to be physically realistic and adhere to the desired potential energy characteristics. This approach provides flexibility in generation—either directly conditioning on a target energy or refining latent samples post hoc via gradient-based optimization.

\subsection{Dimension Choice and Scalability Discussion}
\label{dimension check and scalability}

The choice of latent dimension is critical (Fig. \ref{fig:dim_model-mse}): energy-prediction and edge-reconstruction errors decrease sharply as the latent dimension increases after $dim = 10$, before plateauing in the 16–32 range. A latent code in this latter range effectively captures most structural variability without superfluous model capacity. Furthermore, co-training the decoder with a differentiable energy head invests the latent manifold with thermodynamic significance. By using gradient-steering to guide latent codes into specific energy windows, \textsc{GlassVAE} can synthesize configurations of predetermined stability. This allows for the rapid exploration of low-energy basins, offering an alternative to computationally intensive molecular dynamics sampling.

Although randomly picked snapshots from a single trajectory are structural different, testing on held‑out data alone may not fully demonstrate \textsc{GlassVAE}'s scalability. To address this, we evaluated \textsc{GlassVAE} on two independent Cu$_{50}$Zr$_{50}$ metallic‑glass trajectories at 720K and 960K, both initiated from distinct starting configurations. As shown in Fig.~\ref{fig:scalarable_energy}, the model’s energy predictions for these external trajectories still show accurate prediction patterns as original test set, confirming \textsc{GlassVAE}’s ability to generalize to simulations with markedly different structural evolution. We observe a modest accuracy drop on the new testing trajectories, which likely reflects the limited structural diversity of our training set—only five trajectories at distinct temperatures. Due to computational constraints, we restricted our experiments to these five temperature trajectories to demonstrate crystals or SMILES for molecules. To bridge this gap, we introduce \textsc{GlassVAE}’s capabilities. Incorporating additional trajectories and a wider range of conditions in future work should further enhance the model’s robustness.

\section{Conclusion and Discussion}
Generative modeling of glassy materials has long been hindered by their disordered nature ant the lack of long‑range periodicity, which precludes compact encodings such as CIF for crystals or SMILES for molecules. To bridge this gap, we introduce \textsc{GlassVAE}: a permutation‑, translation‑, and rotation‑invariant graph variational autoencoder with hierarchical latent spaces. The multi‑level embedding employed in this model is essential because a single latent space rarely offers sufficient flexibility to capture both the geometric complexity and property variability of glassy systems, making it challenging to achieve high‐fidelity structure reconstruction and accurate energy prediction simultaneously (Table \ref{tab:main_results}). We further bolster \textsc{GlassVAE} with physics‐informed losses—energy regression, radial distribution function matching, and positional consistency—to ground its generative capabilities in physical reality.

Evaluated on a 108-atom Cu$_{50}$Zr$_{50}$ dataset, \textsc{GlassVAE} demonstrates high fidelity, reconstructing atomic identities and pairwise distances with an $R^2 \sim 0.99$. It also predicts potential energy with a Root Mean Square Error (RMSE) of 0.32 eV/atom for glass configurations over a wide thermodynamic stability. Beyond precise reconstructions, \textsc{GlassVAE} facilitates both unconditional sampling and energy-conditioned refinement. This capability produces new atomic configurations whose energies and radial-distribution functions are consistent with simulation data, underscoring how physics-based regularizers significantly enrich and physically ground the latent space.

To our knowledge, \textsc{GlassVAE} is the first architecture for inorganic glasses to unify structure generation and property prediction within a single, thermodynamically meaningful latent space. In addition, we also demonstrate \textsc{GlassVAE}’s scalability to new Cu$_{50}$Zr$_{50}$ trajectories initialized from different configurations. While current benchmarks use limited data from a 108‑atom system to showcase the initial promise, future work can extend to larger, more complex glasses, disentangle latent factors (e.g., composition, local strain), incorporate additional thermodynamic observables (free‑energy proxies, elastic constants), and integrate active‑learning loops with simulations to further enhance fidelity and guide sampling in high‑dimensional configuration space. Overall, we believe our approach paves the way for accelerated discovery of disordered materials by enabling the simultaneous mapping of energy landscapes and the proposal of candidate structures for further computational or experimental investigations.

\newpage

\setcounter{figure}{0}

\renewcommand{\figurename}{Figure}
\renewcommand{\thefigure}{S\arabic{figure}}

\section{Appendices}

\begin{figure}[h]
    \centering
    \includegraphics[width=0.8\linewidth]{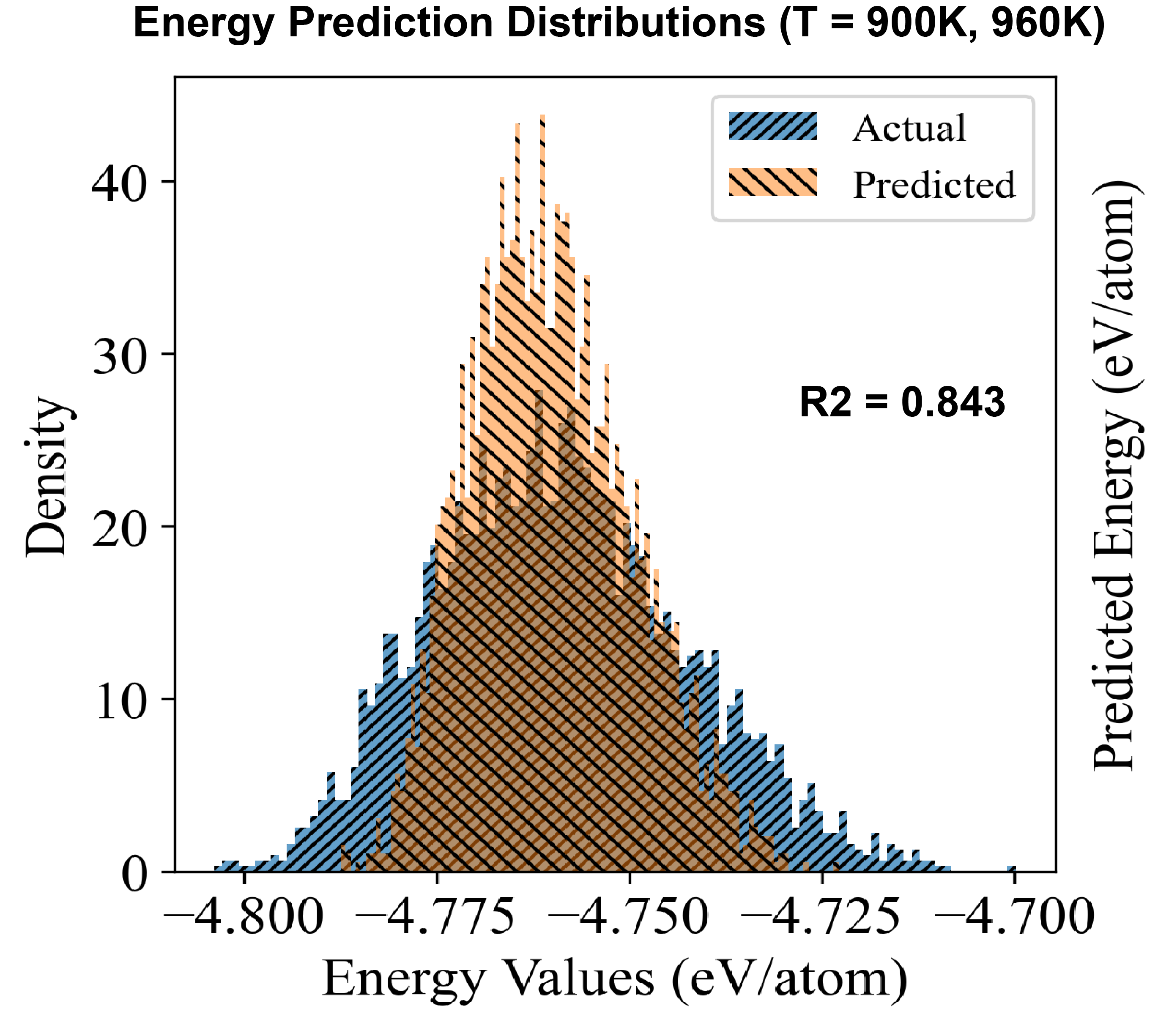}
    \caption{GlassVAE energy prediction performance on a subgroup (900K and 960K) }
    \label{fig:subgroup energyl}
\end{figure}
\begin{figure}[h]
    \centering
    \includegraphics[width=0.8\linewidth]{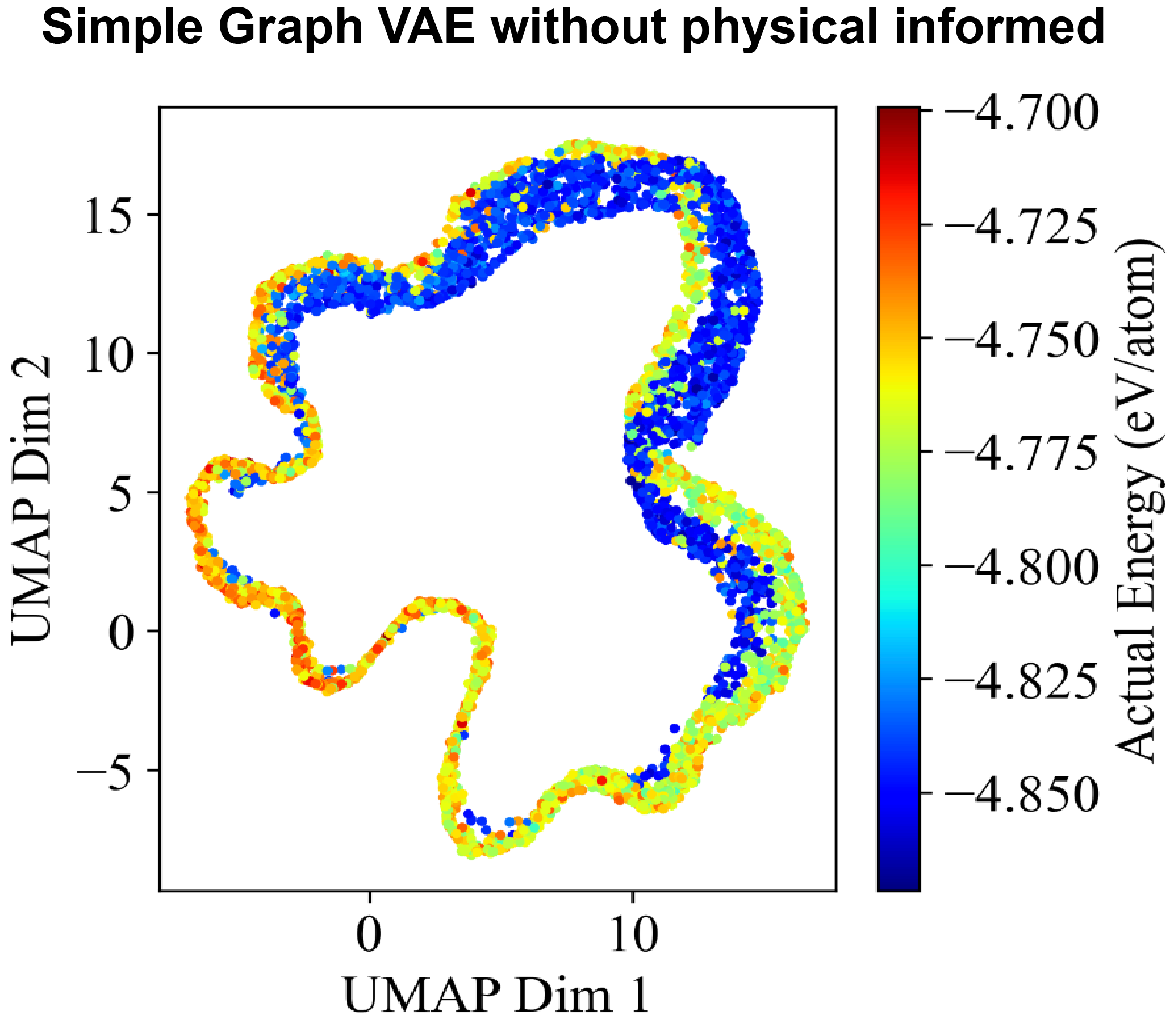}
    \caption{Latent visualization of Graph VAE without physical regularizer}
    \label{fig:simpleVAE w/o Physical}
\end{figure}
\begin{figure*}[h]
    \centering
    \includegraphics[width=0.8\linewidth]{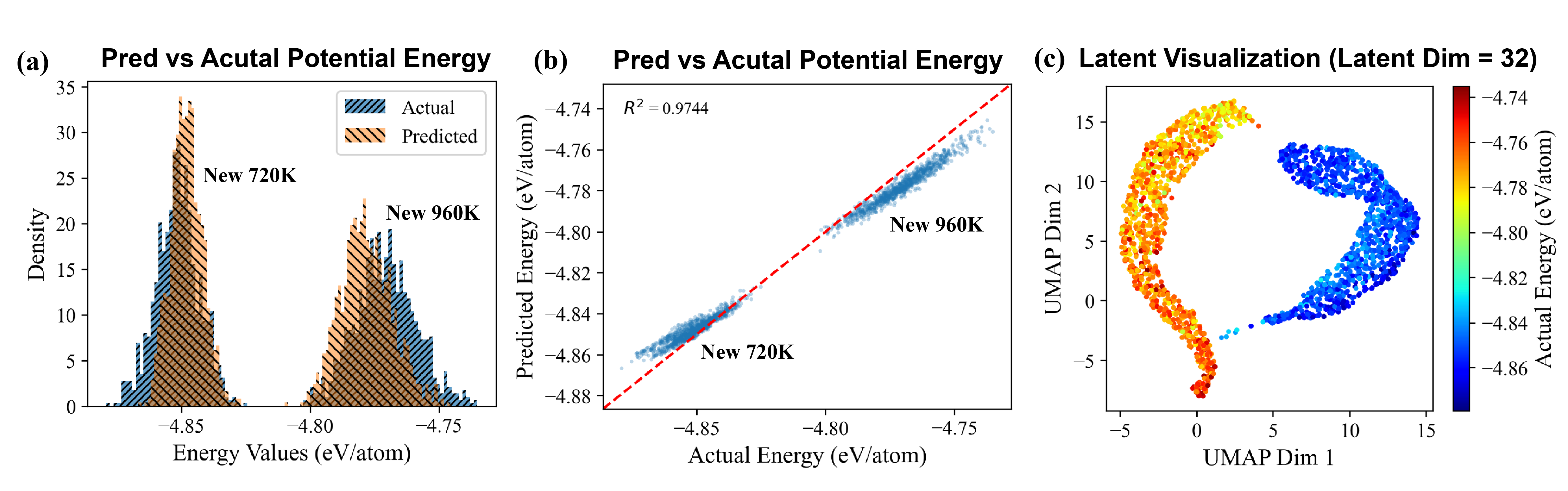}
    \caption{Energy Predictions and Latent Space Visualization of New simulations under 720K and 960K. (a-b) Predicted vs actually potential
energy; (c) visualization of graph latent spaces via UMAP projection}
    \label{fig:scalarable_energy}
\end{figure*}

\begin{figure*}
    \centering
    \includegraphics[width=0.8\linewidth]{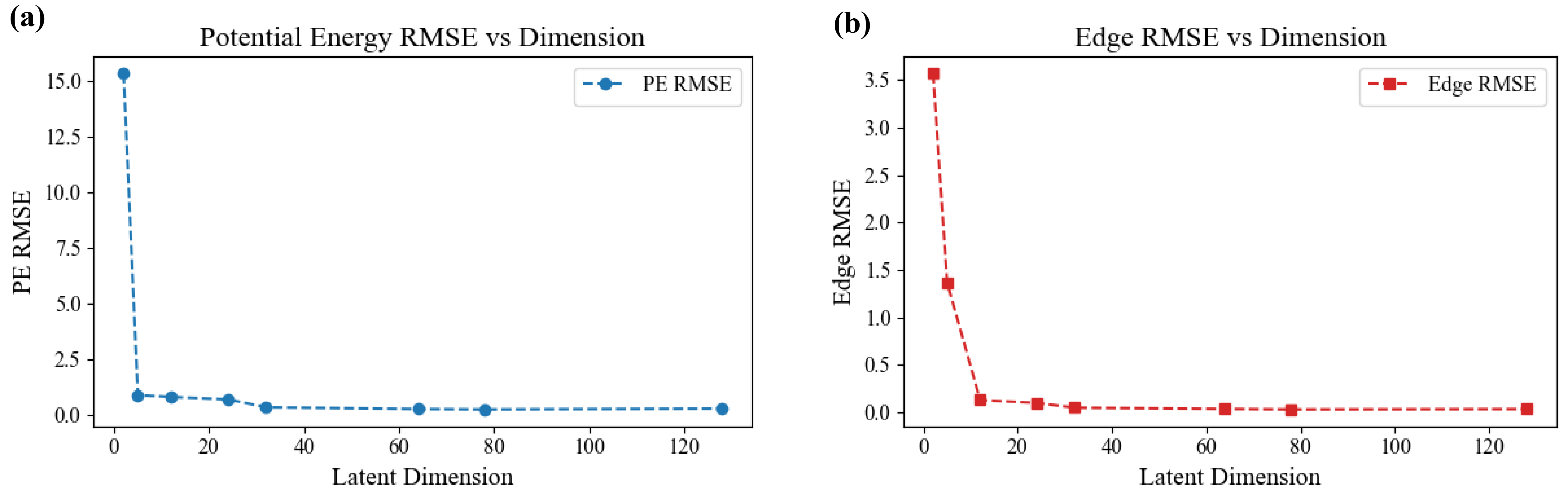}
    \caption{Model performance vs Latent Dimension (a) Potential energy prediction RMSE vs Latent Dimension (b) Edge attribution (pairwise distances) RMSE vs Latent Dimension}
    \label{fig:dim_model-mse}
\end{figure*}

\newpage

\subsection{Algorithm}
\label{Algo}

\begin{algorithm}[H]
\caption{Physical‑Regularized Graph VAE (concise)}
\label{alg:graph_vae_short}

{Atomic configuration $(\mathbf R,\mathbf t,E)$ with $N$ atoms}
{Latent $(\mu,\log\sigma^2)$; reconstructions $\hat{\mathbf R},\hat{Y},\hat{\mathbf e}$; predicted $\hat E$}

\begin{enumerate}
  \item \textbf{Node encoder}\,:\\
        \hspace*{1em}Init node embeddings $\rightarrow$ $L$ message‑passing layers $\rightarrow$ global add‑pool\\
        \hspace*{1em}$\rightarrow$ MLPs $\rightarrow (\mu,\log\sigma^2)$
        
  \item \textbf{Edge encoder}\,:\\
        \hspace*{1em}$L_e$ residual MLP blocks on edge attrs $\rightarrow$ sum to $\mathbf e_{\text{global}}$
        
  \item \textbf{Latent sampling}\,:\\
        \hspace*{1em}$\mathbf z = \mu + \exp(0.5\log\sigma^2)\odot\epsilon,\quad \epsilon\sim\mathcal N(0,I)$

    \item \textbf{Decoder}\
    \begin{itemize}
      \item Node types $\; \hat Y = \mathrm{softmax}(W_n\mathbf z)$
      \item Positions $\; \hat{\mathbf R} = \mathbf R^{(0)} + f_\Delta(\mathbf z)$ 
      \item Edge attrs $\; \hat{\mathbf e}_{ij}=f_e(\mathbf z,\hat{\mathbf r}_j-\hat{\mathbf r}_i)$
    \end{itemize}
\item \textbf{Energy:}

\hspace*{1em}$\; \hat E=f_{\text{energy}}(\mu\Vert\mathbf e_{\text{global}})$

  \item \textbf{Loss}\,:  
  \begin{align*}
      \mathcal{L} = \; &\alpha_{\text{node}} \, \mathcal{L}_{\text{node}} + \alpha_{\text{edge}} \, \mathcal{L}_{\text{edge}} + \alpha_{\text{energy}} \, \mathcal{L}_{\text{energy}} \\
    &+ \beta_{\text{KL}} \, D_{\text{KL}}\bigl(q(z|x)\,\|\,p(z)\bigr) + \alpha_{\text{RDF}} \, \mathcal{L}_{\text{RDF}}
  \end{align*}
  detail of loss terms are list in Sec. \ref{sec:training} 

\end{enumerate}
\end{algorithm}
\newpage

\subsection{Training}
\label{sec:training}

\subsection*{Loss Function}
The overall training objective is a weighted sum of several loss components:
  
\begin{equation}
\begin{aligned}
\mathcal{L} = \; &\alpha_{\text{node}} \, \mathcal{L}_{\text{node}} + \alpha_{\text{edge}} \, \mathcal{L}_{\text{edge}} + \alpha_{\text{energy}} \, \mathcal{L}_{\text{energy}} \\
&+ \beta_{\text{KL}} \, D_{\text{KL}} + \alpha_{\text{RDF}} \, \mathcal{L}_{\text{RDF}}.
\end{aligned}
\label{eq:loss function_SI}
\end{equation}

Node loss is the Binary Cross Entropy (BCE) of predicted and ground truth.

\begin{equation*}
    \mathcal{L}_{\text{node}} = \sum_{i,j} \text{BCE}\left(\hat{x}_{ij}, x_{ij}\right) 
\end{equation*}
Edge loss comprises two components: an MSE loss for the edge distance and an additional term for directional consistency via cosine similarity. Let $\hat{d}_{ij}$ be the predicted distance and $d_{ij}$ the ground-truth distance for edge $(i,j)$. Denote $\hat{\mathbf{a}}_{ij}$ and $\mathbf{a}_{ij}$ as the predicted and true edge feature vectors (excluding the distance component), respectively. Then,
\begin{equation}
\mathcal{L}_{\text{edge}} = \sum_{i,j} \text{MSE}\left(\hat{d}_{ij}, d_{ij}\right) + \lambda \left[1 - \frac{\hat{\mathbf{a}}_{ij} \cdot \mathbf{a}_{ij}}{\|\hat{\mathbf{a}}_{ij}\| \, \|\mathbf{a}_{ij}\|}\right],
\end{equation}
where $\lambda$ is a hyperparameter controlling the contribution of the directional consistency term.

The energy prediction loss is computed as the mean-squared error (MSE) between the predicted energy $\hat{E}$ and the ground-truth energy $E_{\text{true}}$, over a batch of $M$ samples:
\begin{equation}
\mathcal{L}_{\text{energy}} = \frac{1}{M} \sum_{i=1}^{M} \left( \hat{E}_i - E_{\text{true},i} \right)^2.
\end{equation}

The latent space is regularized via the Kullback-Leibler (KL) divergence between the approximate posterior $q(z|x)$ and the prior $p(z)$ (assumed to be standard normal). For a latent vector $z \in \mathbb{R}^{d_z}$ with parameters $\mu \in \mathbb{R}^{d_z}$ and $\sigma \in \mathbb{R}^{d_z}$, this term is expressed as
\begin{equation}
D_{\text{KL}}\bigl(q(z|x) \parallel p(z)\bigr) = -\frac{1}{2} \sum_{k=1}^{d_z} \left( 1 + \log \sigma_k^2 - \mu_k^2 - \exp(\log \sigma_k^2) \right).
\end{equation}

To ensure global structural consistency, we compare soft histograms of pairwise distances computed from the predicted and true atomic positions. Let $\hat{g}_{\text{RDF}}^{(b)}$ and $g_{\text{RDF}}^{(b)}$ denote the RDF histograms for batch sample $b$. Then,
\begin{equation}
\mathcal{L}_{\text{RDF}} = \frac{1}{B} \sum_{b=1}^{B} \left\| \hat{g}_{\text{RDF}}^{(b)} - g_{\text{RDF}}^{(b)} \right\|_2^2,
\end{equation}
where $B$ is the batch size. 

By tuning the weights $\alpha_n$, $\alpha_e$, $\beta_{\rm kl}$, $\alpha_E$, and $\alpha_r$, we balance these objectives to achieve accurate, physically plausible generative performance.

\subsection*{Hyperparameters \& Optimization}
\begin{itemize}
  \item Loss weights: $\alpha_n=1.0,\;\alpha_e=100,\;\beta_{\rm kl}=10^{-4},\;\alpha_E=300,\;\alpha_r=10$.
  \item Optimizer: Adam with gradient clipping $\|\nabla\|\le1.0$.
  \item Box size for RDF: $L = \mathrm{BOX\_SIZE}$.

\end{itemize}

\textbf{Computational Resources}

All experiments and model training were run on the High Throughput Computing (HTC) cluster, using NVIDIA A100‑SXM4‑80GB GPUs. The compute nodes run LINUX with CUDA 11.4, Python 3.9, and PyTorch 2.0.1.

\textbf{MD Experiment Settings}

The model system used in this work is Cu50Zr50, an excellent metallic glass-former, with atomic interactions described using the embedded atom method (EAM) potential. The system was first equilibrated in the liquid state at 2000 K for 2 ns, then quenched to 1000 K in 50 K intervals at a cooling rate of 100 K per 6 ns and further cooled down in 20 K intervals at the rate of 100 K per 60 ns, all using the NPT ensemble. The simulations consisted of 108 atoms, and an MD time step of 1 fs is used. Periodic boundary conditions are employed in all three directions.

\newpage
\subsection{Theorem 2 Full Demonstration}
\label{the2_full}
We define a Sobolev space $\mathcal{U} = W^{s, q}(\mathbb{R}^3, \mathbb{R})$ and Bochner space $\mathcal{V} = L^p(\mathbb{R}^3, \mathbb{R})$ with $1\leq p, q \le \infty$, $s\geq 0$, and norms $||\cdot||_{\mathcal{U}}$ and $||\cdot||_{\mathcal{V}}$, respectively. The physical information of the system is encoded by the abstract form
\begin{equation}
    \mathcal{A}\boldsymbol{u} = v
\end{equation}
, where $\mathcal{A}: \mathcal{U} \rightarrow \mathcal{V}$ is an arbitrary operator satisfying $||v||_{\mathcal{V}} < \infty$, and $||\mathcal{A} \boldsymbol{u}||_{\mathcal{V}} < \infty, \forall \,  ||\boldsymbol{u}||_{\mathcal{U}} < \infty$. As a concrete example of our setting, $\boldsymbol{u}(\cdot; \theta)$ is a neural network parametrized by $\theta$. $\mathcal{A}$ can be the operator that calculates fitted RDF by our neural networks, and $v$ encodes our priori in RDF from the physical law.

\textbf{(Assumption) Regularities of $\mathcal{A}$}
\label{assumption: operator_regularity}

 For any $\boldsymbol{u}, \boldsymbol{u}'\in \mathcal{U}$, 
\begin{equation}\label{eq: operator_regularity}
    ||\boldsymbol{u} - \boldsymbol{u}'||_{\mathcal{U}} \leq C^{(1)} ||\mathcal{A}\boldsymbol{u} - \mathcal{A}\boldsymbol{u}'||_{\mathcal{V}}
\end{equation}

\textbf{(Assumption) Quadrature error}
\label{assumption: quadrature_error}

For an arbitrary function $g \in \mathcal{V}$, we define its integral $\Bar{g} := \int g(\boldsymbol{u})d\boldsymbol{u}$, and the corresponding quadrature approximation
\begin{equation}
    \Bar{g}_N := \sum_{i=1}^N w_ig(\boldsymbol{u}_i)
\end{equation}
, where $\boldsymbol{u}_i$ is the quadrature point with its weights $w_i$. For low-dimensional space, it is safe to assume that 
\begin{equation}\label{eq: quadrature_error}
    |\Bar{g} - \Bar{g}_N| \leq C^{(2)}N^{-\alpha}
\end{equation}
, for some $\alpha > 0$.

The optimal parameter $\theta^*$ is obtained by minimizing the training error
\begin{equation}\label{eq: nn_train_error}
\begin{aligned}
\mathcal{E}_{\text{train}} &:= ||\mathcal{A}\boldsymbol{u}(\cdot; \theta) - v||^2_{L^2} = \int_{\mathbb{R}^3} (\mathcal{A}\boldsymbol{u}(\boldsymbol{x}; \theta) - v)^2\, dx\\
& \approx \sum_{i=1}^N w_i (\mathcal{A}\boldsymbol{u}(\boldsymbol{x}_i, \theta) - v)^2
\end{aligned}
\end{equation}
, where $\{\boldsymbol{x}_i\}_{i=1}^N$ are the sample points in the training set $\mathcal{T}$, with $\{w_i\}_{i=1}^N$ the quadrature weights. We also define the test error
\begin{equation}\label{eq: nn_test_error}
    \mathcal{E}_{\text{test}} := ||\boldsymbol{u}(\cdot; \theta) - \boldsymbol{u}(\cdot; \theta^*)||_{\mathcal{U}}
\end{equation}
The following theorem and its corollary provide an upper bound for the test error in terms of training and validation error.

\begin{theorem}
Suppose the operator $\mathcal{A}$ satisfies \eqref{eq: operator_regularity}, the quadrature error satisfies \eqref{eq: quadrature_error}, and the neural networks are trained by minimizing \eqref{eq: nn_train_error}, then 
\begin{equation}
    \mathcal{E}_{\text{test}} \leq C^{(1)}\mathcal{E}_{\text{train}} + C^{(1)} (C^{(2)})^{1/2} N^{-\alpha/2}
\end{equation}
\end{theorem}

\begin{equation}
\begin{aligned}
\mathcal{E}_{\text{test}} & = ||\boldsymbol{u}(\cdot; \theta) - \boldsymbol{u}(\cdot; \theta^*)||_{\mathcal{U}} \leq C^{(1)} ||\mathcal{A}\boldsymbol{u} - \mathcal{A}\boldsymbol{u}'||_{\mathcal{V}}\\
& \leq C^{(1)}\left(\sum_{i=1}^N w_i (\mathcal{A}\boldsymbol{u}(\boldsymbol{x}_i, \theta) - v)^2  + C^{(2)}N^{-\alpha}\right)^{1/2} \\
& \leq C^{(1)}\mathcal{E}_{\text{train}} + C^{(1)}(C^{(2)})^{1/2}N^{-\alpha/2}
\end{aligned}
\end{equation}

Suppose the operator $\mathcal{A}$ satisfies \eqref{eq: operator_regularity}, the quadrature error satisfies \eqref{eq: quadrature_error}, and the neural networks are trained by minimizing \eqref{eq: nn_train_error}. Furthermore, the validation points and test points are sampled from the same distribution as the training samples $\{\boldsymbol{x}_i\}_{i=1}^N$ in the training set. Then, 
\begin{equation}
\Bar{\mathcal{E}}_{\text{test}} \leq C^{(1)} \left( \Bar{\mathcal{E}}_{\text{train}} + \mathcal{E}_{\text{vg}} + \frac{1}{\sqrt{N}} \sqrt{\hat{\sigma}^2(\theta^*)}\right),
\end{equation}
where $\Bar{\mathcal{E}}_{\text{train}}$ and $\Bar{\mathcal{E}}_{\text{test}}$ is the average train error and test errors across training and test data sets during different training processes, defined as
\begin{equation}
\begin{aligned}
\Bar{\mathcal{E}}_{\text{train}} &= \int_{\mathbb{R}^{3N}} \mathcal{E}_{\text{train}}(\boldsymbol{x}_{1\cdots N}^\mathcal{T})d\mathcal{T}, \quad \Bar{\mathcal{E}}_{\text{test}} &= \int_{\mathbb{R}^{3N}} \mathcal{E}_{\text{test}}(\boldsymbol{x}_{1\cdots N}^\mathcal{T})d\mathcal{T}''
\end{aligned}
\end{equation}
, with $\int \cdot d\mathcal{T}$ and $\int \cdot d\mathcal{T}''$ represents aggregation across different training data sets $\mathcal{T}$ and test data sets $\mathcal{T}''$, respectively. $\Bar{\mathcal{E}}_{\text{valid}}$ represents the expected difference between the training samples and validation samples, defined as
\begin{equation}
    \Bar{\mathcal{E}}_{\text{valid}} = \frac{1}{N} \int_{\mathbb{D}^{3N}} \sum_{j=1} |\mathcal{A}\boldsymbol{u}(\boldsymbol{x}_{1\cdots N}^\mathcal{T}; \theta) - \mathcal{A}\boldsymbol{u}(\boldsymbol{x}_{1\cdots N}^{\mathcal{T}}; \theta^*)| d\mathcal{T}'
\end{equation}
, where  $\int \cdot d\mathcal{T}'$ represents sum across different validation sample sets. $\mathcal{E}_{vg} := \mathbb{E}|\Bar{\mathcal{E}}_{\text{train}} - \Bar{\mathcal{E}}_{\text{test}}|$ is the validation gap between training and validation data set. $\hat{\sigma}^2(\theta^*)$ is the variance of $\theta^*$ obtained by multiple training results,
\begin{equation}
\begin{aligned}
\hat{\sigma}^2(\theta^*) & = \mathbb{E} \Biggl( \int_{\mathbb{R}^{3N}} |\mathcal{A}\boldsymbol{u}(\boldsymbol{x}_{1\cdots N}^\mathcal{T}; \theta) - \mathcal{A}\boldsymbol{u}(\boldsymbol{x}_1^N; \theta^*)|\, d\boldsymbol{x}_{1\cdots N}^\mathcal{T} \\
& \quad\quad  - \int_{\mathcal{T}} \int_{\mathbb{D}^{3N}}|\mathcal{A}\boldsymbol{u}(\boldsymbol{x}_{1\cdots N}^\mathcal{T}; \theta) - \mathcal{A}\boldsymbol{u}(\boldsymbol{x}_1^N; \theta^*)| \, d\boldsymbol{x}_{1\cdots N}^{\mathcal{T}} \, d\mathcal{T} \Biggr)^2.
\end{aligned}
\end{equation}

\begin{equation}
\begin{aligned}
\Bar{\mathcal{E}}_{\text{test}} & = \mathbb{E}||\boldsymbol{u}(\cdot; \theta) - \boldsymbol{u}(\cdot; \theta^*)||_{\mathcal{U}} \leq C^{(1)} \mathbb{E}  ||\mathcal{A}\boldsymbol{u} - \mathcal{A}\boldsymbol{u}'||_{\mathcal{V}}\\
& \leq C^{(1)} \left( \mathbb{E}\left(\mathbb{E}(||\mathcal{A}\boldsymbol{u} - \mathcal{A}\boldsymbol{u}'||_{\mathcal{V}})|\mathcal{T}\right) + \Bar{\mathcal{E}}_{\text{valid}} - \Bar{\mathcal{E}}_{\text{valid}} + \Bar{\mathcal{E}}_{\text{test}} - \Bar{\mathcal{E}}_{\text{test}}\right)\\
& \leq C^{(1)} \left( \Bar{\mathcal{E}}_{\text{train}} + \mathcal{E}_{\text{vg}} + \frac{1}{\sqrt{N}} \sqrt{ \mathbb{E} \left( \left| \mathbb{E} \left(||\mathcal{A}\boldsymbol{u} - \mathcal{A}\boldsymbol{u}'||_{\mathcal{V}}|\mathcal{T} \right) - \bar{\mathcal{E}}_V \right|^2 \right) }
\right)
\end{aligned}
\end{equation}
, and $\hat{\sigma}^2(\theta^*)$ estimates $\mathbb{E} \left( \left| \mathbb{E} \left(||\mathcal{A}\boldsymbol{u} - \mathcal{A}\boldsymbol{u}'||_{\mathcal{V}}|\mathcal{T} \right) - \bar{\mathcal{E}}_V \right|^2 \right)$.

\subsection{Data Pre-processing}
\label{dataprocess}

Before model training, we convert raw simulation outputs into a graph‐based dataset in four stages. First, we parse LAMMPS files to extract atomic coordinates and element labels, and we read corresponding potential energies from the EAM output and get \(\{\mathbf{r}_i, t_i, E\}\) . To ensure balanced sampling across thermodynamic states, we randomly select up to 5000 snapshots per temperature. Next, we normalize the energy values to a 0–100 scale, avoiding division by zero when the range is degenerate. Each configuration is then transformed into a PyTorch‐Geometric Data object: atom types become one‐hot node features, interatomic displacements (with periodic boundary corrections) and distances form edge attributes, and we store true positions for later radial‐distribution‐function regularization. 

\textbf{Periodic-boundary displacements:}

    For each pair \((i,j)\),
    \[
      \Delta \mathbf{r}_{ij}
      = \bigl(\mathbf{r}_i - \mathbf{r}_j + \tfrac{L}{2}\bigr)\bmod L \;-\; \tfrac{L}{2},
      \quad
      d_{ij} = \|\Delta \mathbf{r}_{ij}\|_2,
    \]
    where \(L\) is the box length.
    
Finally, we assemble these graphs into an 80/20 train–test split and wrap them in DataLoaders (batch size = 64, also shuffled for training) to feed into our GlassVAE pipeline.

\section*{Acknowledgments}
This research was primarily supported by NSF through the University of Wisconsin Materials Research Science and Engineering Center (DMR-2309000)

\newpage

\bibliography{GLASSVAE}

@book{allen2017computer,
  title={Computer simulation of liquids},
  author={Allen, Michael P and Tildesley, Dominic J},
  year={2017},
  publisher={Oxford university press}
}

@article{bamerMolecularMechanicsDisordered2023,
  title = {Molecular {{Mechanics}} of {{Disordered Solids}}},
  author = {Bamer, Franz and Ebrahem, Firaz and Markert, Bernd and Stamm, Benjamin},
  year = {2023},
  month = apr,
  journal = {Arch Computat Methods Eng},
  volume = {30},
  number = {3},
  pages = {2105--2180},
  issn = {1886-1784},
  doi = {10.1007/s11831-022-09861-1},
  urldate = {2025-05-12},
  langid = {english}
}

@article{court3DInorganicCrystal2020,
  title = {3-{{D Inorganic Crystal Structure Generation}} and {{Property Prediction}} via {{Representation Learning}}},
  author = {Court, Callum J. and Yildirim, Batuhan and Jain, Apoorv and Cole, Jacqueline M.},
  year = {2020},
  month = oct,
  journal = {J. Chem. Inf. Model.},
  volume = {60},
  number = {10},
  pages = {4518--4535},
  publisher = {American Chemical Society},
  issn = {1549-9596},
  doi = {10.1021/acs.jcim.0c00464},
  urldate = {2025-05-12}
}

@misc{flam-shepherdLanguageModelsCan2023,
  title = {Language Models Can Generate Molecules, Materials, and Protein Binding Sites Directly in Three Dimensions as {{XYZ}}, {{CIF}}, and {{PDB}} Files},
  author = {{Flam-Shepherd}, Daniel and {Aspuru-Guzik}, Al{\'a}n},
  year = {2023},
  month = may,
  number = {arXiv:2305.05708},
  eprint = {2305.05708},
  primaryclass = {cs},
  publisher = {arXiv},
  doi = {10.48550/arXiv.2305.05708},
  urldate = {2025-05-12},
  archiveprefix = {arXiv}
}

@article{jungRoadmapMachineLearning2025,
  title = {Roadmap on Machine Learning Glassy Dynamics},
  author = {Jung, Gerhard and Alkemade, Rinske M. and Bapst, Victor and Coslovich, Daniele and Filion, Laura and Landes, Fran{\c c}ois P. and Liu, Andrea and Pezzicoli, Francesco Saverio and Shiba, Hayato and Volpe, Giovanni and Zamponi, Francesco and Berthier, Ludovic and Biroli, Giulio},
  year = {2025},
  month = jan,
  journal = {Nat Rev Phys},
  volume = {7},
  number = {2},
  eprint = {2311.14752},
  primaryclass = {cond-mat},
  pages = {91--104},
  issn = {2522-5820},
  doi = {10.1038/s42254-024-00791-4},
  urldate = {2025-03-14},
  archiveprefix = {arXiv},
  langid = {english}
}

@inproceedings{klushynLearningHierarchicalPriors2019,
  title = {Learning {{Hierarchical Priors}} in {{VAEs}}},
  booktitle = {Advances in {{Neural Information Processing Systems}}},
  author = {Klushyn, Alexej and Chen, Nutan and Kurle, Richard and Cseke, Botond and {van der Smagt}, Patrick},
  year = {2019},
  volume = {32},
  publisher = {Curran Associates, Inc.},
  urldate = {2025-05-12}
}

@article{liGenerativeDesignCrystal2025,
  title = {Generative Design of Crystal Structures by Point Cloud Representations and Diffusion Model},
  author = {Li, Zhelin and Mrad, Rami and Jiao, Runxian and Huang, Guan and Shan, Jun and Chu, Shibing and Chen, Yuanping},
  year = {2025},
  month = jan,
  journal = {iScience},
  volume = {28},
  number = {1},
  pages = {111659},
  issn = {25890042},
  doi = {10.1016/j.isci.2024.111659},
  urldate = {2025-03-14},
  langid = {english}
}

@misc{liPredictingInterpretingEnergy2024,
  title = {Predicting and {{Interpreting Energy Barriers}} of {{Metallic Glasses}} with {{Graph Neural Networks}}},
  author = {Li, Haoyu and Zhang, Shichang and Tang, Longwen and Bauchy, Mathieu and Sun, Yizhou},
  year = {2024},
  month = sep,
  number = {arXiv:2401.08627},
  eprint = {2401.08627},
  primaryclass = {cond-mat},
  publisher = {arXiv},
  doi = {10.48550/arXiv.2401.08627},
  urldate = {2025-03-14},
  archiveprefix = {arXiv},
  langid = {english}
}

@article{longConstrainedCrystalsDeep2021,
  title = {Constrained Crystals Deep Convolutional Generative Adversarial Network for the Inverse Design of Crystal Structures},
  author = {Long, Teng and Fortunato, Nuno M. and Opahle, Ingo and Zhang, Yixuan and Samathrakis, Ilias and Shen, Chen and Gutfleisch, Oliver and Zhang, Hongbin},
  year = {2021},
  month = may,
  journal = {npj Comput Mater},
  volume = {7},
  number = {1},
  pages = {1--7},
  publisher = {Nature Publishing Group},
  issn = {2057-3960},
  doi = {10.1038/s41524-021-00526-4},
  urldate = {2025-05-12},
  copyright = {2021 The Author(s)},
  langid = {english}
}

@misc{nouiraCrystalGANLearningDiscover2019,
  title = {{{CrystalGAN}}: {{Learning}} to {{Discover Crystallographic Structures}} with {{Generative Adversarial Networks}}},
  shorttitle = {{{CrystalGAN}}},
  author = {Nouira, Asma and Sokolovska, Nataliya and Crivello, Jean-Claude},
  year = {2019},
  month = may,
  number = {arXiv:1810.11203},
  eprint = {1810.11203},
  primaryclass = {cs},
  publisher = {arXiv},
  doi = {10.48550/arXiv.1810.11203},
  urldate = {2025-05-12},
  archiveprefix = {arXiv}
}

@inproceedings{vahdatNVAEDeepHierarchical2020,
  title = {{{NVAE}}: {{A Deep Hierarchical Variational Autoencoder}}},
  shorttitle = {{{NVAE}}},
  booktitle = {Advances in {{Neural Information Processing Systems}}},
  author = {Vahdat, Arash and Kautz, Jan},
  year = {2020},
  volume = {33},
  pages = {19667--19679},
  publisher = {Curran Associates, Inc.},
  urldate = {2025-05-12}
}

@article{xuGenerativeAdversarialNetworks2023,
  title = {A {{Generative Adversarial Networks}} ({{GAN}}) Based Efficient Sampling Method for Inverse Design of Metallic Glasses},
  author = {Xu, Xiang and Hu, Jingyi},
  year = {2023},
  month = aug,
  journal = {Journal of Non-Crystalline Solids},
  volume = {613},
  pages = {122378},
  issn = {00223093},
  doi = {10.1016/j.jnoncrysol.2023.122378},
  urldate = {2025-03-14},
  langid = {english}
}

@misc{yingHierarchicalGraphRepresentation2019,
  title = {Hierarchical {{Graph Representation Learning}} with {{Differentiable Pooling}}},
  author = {Ying, Rex and You, Jiaxuan and Morris, Christopher and Ren, Xiang and Hamilton, William L. and Leskovec, Jure},
  year = {2019},
  month = feb,
  number = {arXiv:1806.08804},
  eprint = {1806.08804},
  primaryclass = {cs},
  publisher = {arXiv},
  doi = {10.48550/arXiv.1806.08804},
  urldate = {2025-05-12},
  archiveprefix = {arXiv}
}

@misc{zeniMatterGenGenerativeModel2024,
  title = {{{MatterGen}}: A Generative Model for Inorganic Materials Design},
  shorttitle = {{{MatterGen}}},
  author = {Zeni, Claudio and Pinsler, Robert and Z{\"u}gner, Daniel and Fowler, Andrew and Horton, Matthew and Fu, Xiang and Shysheya, Sasha and Crabb{\'e}, Jonathan and Sun, Lixin and Smith, Jake and Nguyen, Bichlien and Schulz, Hannes and Lewis, Sarah and Huang, Chin-Wei and Lu, Ziheng and Zhou, Yichi and Yang, Han and Hao, Hongxia and Li, Jielan and Tomioka, Ryota and Xie, Tian},
  year = {2024},
  month = jan,
  number = {arXiv:2312.03687},
  eprint = {2312.03687},
  primaryclass = {cond-mat},
  publisher = {arXiv},
  doi = {10.48550/arXiv.2312.03687},
  urldate = {2025-05-12},
  archiveprefix = {arXiv}
}

@misc{oboyle_deepsmiles_2018,
    title = {{DeepSMILES}: {An} {Adaptation} of {SMILES} for {Use} in {Machine}-{Learning} of {Chemical} {Structures}},
    shorttitle = {{DeepSMILES}},
    url = {https://chemrxiv.org/engage/chemrxiv/article-details/60c73ed6567dfe7e5fec388d},
    doi = {10.26434/chemrxiv.7097960.v1},
    language = {en},
    urldate = {2025-05-12},
    publisher = {ChemRxiv},
    author = {O'Boyle, Noel and Dalke, Andrew},
    month = sep,
    year = {2018},
}

@article{skinnider_invalid_2024,
    title = {Invalid {SMILES} are beneficial rather than detrimental to chemical language models},
    volume = {6},
    copyright = {2024 The Author(s)},
    issn = {2522-5839},
    url = {https://www.nature.com/articles/s42256-024-00821-x},
    doi = {10.1038/s42256-024-00821-x},
    language = {en},
    number = {4},
    urldate = {2025-05-12},
    journal = {Nature Machine Intelligence},
    author = {Skinnider, Michael A.},
    month = apr,
    year = {2024},
    note = {Publisher: Nature Publishing Group},
    pages = {437--448},
}

@article{weininger_smiles_1988,
    title = {{SMILES}, a chemical language and information system. 1. {Introduction} to methodology and encoding rules},
    volume = {28},
    issn = {0095-2338},
    url = {https://doi.org/10.1021/ci00057a005},
    doi = {10.1021/ci00057a005},
    number = {1},
    urldate = {2025-05-12},
    journal = {Journal of Chemical Information and Computer Sciences},
    author = {Weininger, David},
    month = feb,
    year = {1988},
    note = {Publisher: American Chemical Society},
    pages = {31--36},
}

@article{de_comparing_2016,
    title = {Comparing molecules and solids across structural and alchemical space},
    volume = {18},
    issn = {1463-9076, 1463-9084},
    url = {http://arxiv.org/abs/1601.04077},
    doi = {10.1039/C6CP00415F},
    number = {20},
    urldate = {2025-05-12},
    journal = {Physical Chemistry Chemical Physics},
    author = {De, Sandip and Bartók, Albert P. and Csányi, Gábor and Ceriotti, Michele},
    year = {2016},
    note = {arXiv:1601.04077 [cond-mat]},
    pages = {13754--13769},
}

@article{lin_expanding_2024,
    title = {Expanding density-correlation machine learning representations for anisotropic coarse-grained particles},
    volume = {161},
    issn = {0021-9606},
    url = {https://doi.org/10.1063/5.0210910},
    doi = {10.1063/5.0210910},
    number = {7},
    urldate = {2025-05-12},
    journal = {The Journal of Chemical Physics},
    author = {Lin, Arthur and Huguenin-Dumittan, Kevin K. and Cho, Yong-Cheol and Nigam, Jigyasa and Cersonsky, Rose K.},
    month = aug,
    year = {2024},
    pages = {074112},
}

@article{bartok_representing_2013,
    title = {On representing chemical environments},
    volume = {87},
    issn = {1098-0121, 1550-235X},
    url = {http://arxiv.org/abs/1209.3140},
    doi = {10.1103/PhysRevB.87.184115},
    number = {18},
    urldate = {2025-05-12},
    journal = {Physical Review B},
    author = {Bartók, Albert P. and Kondor, Risi and Csányi, Gábor},
    month = may,
    year = {2013},
    note = {arXiv:1209.3140 [physics]},
    pages = {184115},
}

@article{mudassir_systematic_2022,
    title = {Systematic {Identification} of {Atom}-{Centered} {Symmetry} {Functions} for the {Development} of {Neural} {Network} {Potentials}},
    volume = {126},
    issn = {1089-5639},
    url = {https://doi.org/10.1021/acs.jpca.2c04508},
    doi = {10.1021/acs.jpca.2c04508},
    number = {44},
    urldate = {2025-05-12},
    journal = {The Journal of Physical Chemistry A},
    author = {Mudassir, Mohammed Wasay and Goverapet Srinivasan, Sriram and Mynam, Mahesh and Rai, Beena},
    month = nov,
    year = {2022},
    note = {Publisher: American Chemical Society},
    pages = {8337--8347},
}

@article{schroers_thermoplastic_2011,
    title = {Thermoplastic blow molding of metals},
    volume = {14},
    issn = {1369-7021},
    url = {https://www.sciencedirect.com/science/article/pii/S1369702111700189},
    doi = {10.1016/S1369-7021(11)70018-9},
    number = {1},
    urldate = {2025-05-12},
    journal = {Materials Today},
    author = {Schroers, Jan and Hodges, Thomas M. and Kumar, Golden and Raman, Hari and Barnes, Anthony J. and Pham, Quoc and Waniuk, Theodore A.},
    month = jan,
    year = {2011},
    pages = {14--19},
}

@article{trexler_mechanical_2010,
    title = {Mechanical properties of bulk metallic glasses},
    volume = {55},
    issn = {0079-6425},
    url = {https://www.sciencedirect.com/science/article/pii/S0079642510000277},
    doi = {10.1016/j.pmatsci.2010.04.002},
    number = {8},
    urldate = {2025-05-12},
    journal = {Progress in Materials Science},
    author = {Trexler, Morgana Martin and Thadhani, Naresh N.},
    month = nov,
    year = {2010},
    pages = {759--839},
}

@incollection{bansal_chapter_1986,
    address = {San Diego},
    title = {Chapter 1 - {Introduction}},
    isbn = {978-0-08-052376-7},
    url = {https://www.sciencedirect.com/science/article/pii/B9780080523767500047},
    urldate = {2025-05-12},
    booktitle = {Handbook of {Glass} {Properties}},
    publisher = {Academic Press},
    author = {Bansal, NAROTTAM P. and Doremus, R. H.},
    editor = {Bansal, NAROTTAM P. and Doremus, R. H.},
    month = jan,
    year = {1986},
    doi = {10.1016/B978-0-08-052376-7.50004-7},
    pages = {1--3},
}

@article{starr_what_2002,
    title = {What do we learn from the local geometry of glass-forming liquids?},
    volume = {89},
    issn = {0031-9007, 1079-7114},
    url = {http://arxiv.org/abs/cond-mat/0206121},
    doi = {10.1103/PhysRevLett.89.125501},
    number = {12},
    urldate = {2025-05-12},
    journal = {Physical Review Letters},
    author = {Starr, Francis W. and Sastry, Srikanth and Douglas, Jack F. and Glotzer, Sharon C.},
    month = aug,
    year = {2002},
    note = {arXiv:cond-mat/0206121},
    pages = {125501},
}

@article{cao_structural_2018,
    title = {Structural and topological nature of plasticity in sheared granular materials},
    volume = {9},
    copyright = {2018 The Author(s)},
    issn = {2041-1723},
    url = {https://www.nature.com/articles/s41467-018-05329-8},
    doi = {10.1038/s41467-018-05329-8},
    language = {en},
    number = {1},
    urldate = {2025-05-12},
    journal = {Nature Communications},
    author = {Cao, Yixin and Li, Jindong and Kou, Binquan and Xia, Chengjie and Li, Zhifeng and Chen, Rongchang and Xie, Honglan and Xiao, Tiqiao and Kob, Walter and Hong, Liang and Zhang, Jie and Wang, Yujie},
    month = jul,
    year = {2018},
    note = {Publisher: Nature Publishing Group},
    pages = {2911},
}

@book{rapaport_art_2004,
    address = {Cambridge},
    edition = {2},
    title = {The {Art} of {Molecular} {Dynamics} {Simulation}},
    isbn = {978-0-521-82568-9},
    url = {https://www.cambridge.org/core/books/art-of-molecular-dynamics-simulation/57D40C5ECE9B7EA17C0E77E7754F5874},
    urldate = {2025-05-12},
    publisher = {Cambridge University Press},
    author = {Rapaport, D. C.},
    year = {2004},
    doi = {10.1017/CBO9780511816581},
}

@article{pauly_transformation-mediated_2010,
    title = {Transformation-mediated ductility in {CuZr}-based bulk metallic glasses},
    volume = {9},
    copyright = {2010 Springer Nature Limited},
    issn = {1476-4660},
    url = {https://www.nature.com/articles/nmat2767},
    doi = {10.1038/nmat2767},
    language = {en},
    number = {6},
    urldate = {2025-05-12},
    journal = {Nature Materials},
    author = {Pauly, S. and Gorantla, S. and Wang, G. and Kühn, U. and Eckert, J.},
    month = jun,
    year = {2010},
    note = {Publisher: Nature Publishing Group},
    pages = {473--477},
}

@article{wang_inverse_2021,
    title = {Inverse design of glass structure with deep graph neural networks},
    volume = {12},
    copyright = {2021 The Author(s)},
    issn = {2041-1723},
    url = {https://www.nature.com/articles/s41467-021-25490-x},
    doi = {10.1038/s41467-021-25490-x},
    language = {en},
    number = {1},
    urldate = {2025-05-12},
    journal = {Nature Communications},
    author = {Wang, Qi and Zhang, Longfei},
    month = sep,
    year = {2021},
    note = {Publisher: Nature Publishing Group},
    pages = {5359},
}

@misc{yoshikawa_graph_2025,
    title = {Graph {Neural} {Network}-based structural classification of glass-forming liquids and its interpretation via {Self}-{Attention} mechanism},
    url = {http://arxiv.org/abs/2505.00993},
    doi = {10.48550/arXiv.2505.00993},
    urldate = {2025-05-12},
    publisher = {arXiv},
    author = {Yoshikawa, Kohei and Yano, Kentaro and Goto, Shota and Kim, Kang and Matubayasi, Nobuyuki},
    month = may,
    year = {2025},
    note = {arXiv:2505.00993 [cond-mat]},
}

@article{bapst_unveiling_2020,
    title = {Unveiling the predictive power of static structure in glassy systems},
    volume = {16},
    copyright = {2020 The Author(s), under exclusive licence to Springer Nature Limited},
    issn = {1745-2481},
    url = {https://www.nature.com/articles/s41567-020-0842-8},
    doi = {10.1038/s41567-020-0842-8},
    language = {en},
    number = {4},
    urldate = {2025-05-12},
    journal = {Nature Physics},
    author = {Bapst, V. and Keck, T. and Grabska-Barwińska, A. and Donner, C. and Cubuk, E. D. and Schoenholz, S. S. and Obika, A. and Nelson, A. W. R. and Back, T. and Hassabis, D. and Kohli, P.},
    month = apr,
    year = {2020},
    note = {Publisher: Nature Publishing Group},
    pages = {448--454},
}

@misc{gilmer_neural_2017,
    title = {Neural {Message} {Passing} for {Quantum} {Chemistry}},
    url = {http://arxiv.org/abs/1704.01212},
    doi = {10.48550/arXiv.1704.01212},
    urldate = {2025-05-12},
    publisher = {arXiv},
    author = {Gilmer, Justin and Schoenholz, Samuel S. and Riley, Patrick F. and Vinyals, Oriol and Dahl, George E.},
    month = jun,
    year = {2017},
    note = {arXiv:1704.01212 [cs]},
}

@misc{satorras_en_2022,
    title = {E(n) {Equivariant} {Graph} {Neural} {Networks}},
    url = {http://arxiv.org/abs/2102.09844},
    doi = {10.48550/arXiv.2102.09844},
    urldate = {2025-05-12},
    publisher = {arXiv},
    author = {Satorras, Victor Garcia and Hoogeboom, Emiel and Welling, Max},
    month = feb,
    year = {2022},
    note = {arXiv:2102.09844 [cs]},
}

@misc{sanchez-gonzalez_learning_2020,
    title = {Learning to {Simulate} {Complex} {Physics} with {Graph} {Networks}},
    url = {http://arxiv.org/abs/2002.09405},
    doi = {10.48550/arXiv.2002.09405},
    urldate = {2025-05-12},
    publisher = {arXiv},
    author = {Sanchez-Gonzalez, Alvaro and Godwin, Jonathan and Pfaff, Tobias and Ying, Rex and Leskovec, Jure and Battaglia, Peter W.},
    month = sep,
    year = {2020},
    note = {arXiv:2002.09405 [cs]},
}

@misc{krykunov_bond_2018,
    title = {Bond type restricted radial distribution functions for accurate machine learning prediction of atomization energies},
    url = {http://arxiv.org/abs/1807.10301},
    doi = {10.48550/arXiv.1807.10301},
    urldate = {2025-05-12},
    publisher = {arXiv},
    author = {Krykunov, Mykhaylo and Woo, Tom K.},
    month = jul,
    year = {2018},
    note = {arXiv:1807.10301 [physics]},
}

@article{watanabe_machine_2024,
    title = {A machine learning potential construction based on radial distribution function sampling},
    volume = {45},
    copyright = {© 2024 The Author(s). Journal of Computational Chemistry published by Wiley Periodicals LLC.},
    issn = {1096-987X},
    url = {https://onlinelibrary.wiley.com/doi/abs/10.1002/jcc.27497},
    doi = {10.1002/jcc.27497},
    language = {en},
    number = {32},
    urldate = {2025-05-12},
    journal = {Journal of Computational Chemistry},
    author = {Watanabe, Natsuki and Hori, Yuta and Sugisawa, Hiroki and Ida, Tomonori and Shoji, Mitsuo and Shigeta, Yasuteru},
    year = {2024},
    note = {\_eprint: https://onlinelibrary.wiley.com/doi/pdf/10.1002/jcc.27497},
    pages = {2949--2958},
}

@misc{nouira_crystalgan_2019,
    title = {{CrystalGAN}: {Learning} to {Discover} {Crystallographic} {Structures} with {Generative} {Adversarial} {Networks}},
    shorttitle = {{CrystalGAN}},
    url = {http://arxiv.org/abs/1810.11203},
    doi = {10.48550/arXiv.1810.11203},
    urldate = {2025-05-12},
    publisher = {arXiv},
    author = {Nouira, Asma and Sokolovska, Nataliya and Crivello, Jean-Claude},
    month = may,
    year = {2019},
    note = {arXiv:1810.11203 [cs]},
}

@article{gomez-bombarelli_automatic_2018,
    title = {Automatic {Chemical} {Design} {Using} a {Data}-{Driven} {Continuous} {Representation} of {Molecules}},
    volume = {4},
    issn = {2374-7943},
    url = {https://doi.org/10.1021/acscentsci.7b00572},
    doi = {10.1021/acscentsci.7b00572},
    number = {2},
    urldate = {2025-05-12},
    journal = {ACS Central Science},
    author = {Gómez-Bombarelli, Rafael and Wei, Jennifer N. and Duvenaud, David and Hernández-Lobato, José Miguel and Sánchez-Lengeling, Benjamín and Sheberla, Dennis and Aguilera-Iparraguirre, Jorge and Hirzel, Timothy D. and Adams, Ryan P. and Aspuru-Guzik, Alán},
    month = feb,
    year = {2018},
    note = {Publisher: American Chemical Society},
    pages = {268--276},
}

@article{luo_deep_2024,
    title = {Deep learning generative model for crystal structure prediction},
    volume = {10},
    copyright = {2024 The Author(s)},
    issn = {2057-3960},
    url = {https://www.nature.com/articles/s41524-024-01443-y},
    doi = {10.1038/s41524-024-01443-y},
    language = {en},
    number = {1},
    urldate = {2025-05-12},
    journal = {npj Computational Materials},
    author = {Luo, Xiaoshan and Wang, Zhenyu and Gao, Pengyue and Lv, Jian and Wang, Yanchao and Chen, Changfeng and Ma, Yanming},
    month = nov,
    year = {2024},
    note = {Publisher: Nature Publishing Group},
    pages = {1--10},
}

@article{wang_compositionally_2021,
    title = {Compositionally restricted attention-based network for materials property predictions},
    volume = {7},
    copyright = {2021 The Author(s)},
    issn = {2057-3960},
    url = {https://www.nature.com/articles/s41524-021-00545-1},
    doi = {10.1038/s41524-021-00545-1},
    language = {en},
    number = {1},
    urldate = {2025-05-12},
    journal = {npj Computational Materials},
    author = {Wang, Anthony Yu-Tung and Kauwe, Steven K. and Murdock, Ryan J. and Sparks, Taylor D.},
    month = may,
    year = {2021},
    note = {Publisher: Nature Publishing Group},
    pages = {1--10},
}

@misc{bao_equivariant_2023,
    title = {Equivariant {Energy}-{Guided} {SDE} for {Inverse} {Molecular} {Design}},
    url = {http://arxiv.org/abs/2209.15408},
    doi = {10.48550/arXiv.2209.15408},
    urldate = {2025-05-12},
    publisher = {arXiv},
    author = {Bao, Fan and Zhao, Min and Hao, Zhongkai and Li, Peiyao and Li, Chongxuan and Zhu, Jun},
    month = mar,
    year = {2023},
    note = {arXiv:2209.15408 [physics]},
}

@article{schleder_dft_2019,
    title = {From {DFT} to machine learning: recent approaches to materials science–a review},
    volume = {2},
    issn = {2515-7639},
    shorttitle = {From {DFT} to machine learning},
    url = {https://dx.doi.org/10.1088/2515-7639/ab084b},
    doi = {10.1088/2515-7639/ab084b},
    language = {en},
    number = {3},
    urldate = {2025-05-12},
    journal = {Journal of Physics: Materials},
    author = {Schleder, Gabriel R and Padilha, Antonio C M and Acosta, Carlos Mera and Costa, Marcio and Fazzio, Adalberto},
    month = may,
    year = {2019},
    note = {Publisher: IOP Publishing},
    pages = {032001},
}

@article{xie_graph_2019,
    title = {Graph dynamical networks for unsupervised learning of atomic scale dynamics in materials},
    volume = {10},
    copyright = {2019 The Author(s)},
    issn = {2041-1723},
    url = {https://www.nature.com/articles/s41467-019-10663-6},
    doi = {10.1038/s41467-019-10663-6},
    language = {en},
    number = {1},
    urldate = {2025-05-12},
    journal = {Nature Communications},
    author = {Xie, Tian and France-Lanord, Arthur and Wang, Yanming and Shao-Horn, Yang and Grossman, Jeffrey C.},
    month = jun,
    year = {2019},
    note = {Publisher: Nature Publishing Group},
    pages = {2667},
}

@article{ren_invertible_2022,
    title = {An invertible crystallographic representation for general inverse design of inorganic crystals with targeted properties},
    volume = {5},
    issn = {25902385},
    url = {http://arxiv.org/abs/2005.07609},
    doi = {10.1016/j.matt.2021.11.032},
    number = {1},
    urldate = {2025-05-12},
    journal = {Matter},
    author = {Ren, Zekun and Tian, Siyu Isaac Parker and Noh, Juhwan and Oviedo, Felipe and Xing, Guangzong and Li, Jiali and Liang, Qiaohao and Zhu, Ruiming and Aberle, Armin G. and Sun, Shijing and Wang, Xiaonan and Liu, Yi and Li, Qianxiao and Jayavelu, Senthilnath and Hippalgaonkar, Kedar and Jung, Yousung and Buonassisi, Tonio},
    month = jan,
    year = {2022},
    note = {arXiv:2005.07609 [physics]},
    pages = {314--335},
}

@misc{hoogeboom_equivariant_2022,
    title = {Equivariant {Diffusion} for {Molecule} {Generation} in {3D}},
    url = {http://arxiv.org/abs/2203.17003},
    doi = {10.48550/arXiv.2203.17003},
    urldate = {2025-05-12},
    publisher = {arXiv},
    author = {Hoogeboom, Emiel and Satorras, Victor Garcia and Vignac, Clément and Welling, Max},
    month = jun,
    year = {2022},
    note = {arXiv:2203.17003 [cs]},
}

@misc{sun_ice_2024,
    title = {Ice phase classification made easy with score-based denoising},
    url = {http://arxiv.org/abs/2405.06599},
    doi = {10.48550/arXiv.2405.06599},
    urldate = {2025-05-12},
    publisher = {arXiv},
    author = {Sun, Hong and Hamel, Sebastien and Hsu, Tim and Sadigh, Babak and Lordi, Vince and Zhou, Fei},
    month = may,
    year = {2024},
    note = {arXiv:2405.06599 [cond-mat]
version: 1},
}

@article{li_inverse_2024,
    title = {Inverse design machine learning model for metallic glasses with good glass-forming ability and properties},
    volume = {135},
    issn = {0021-8979},
    url = {https://doi.org/10.1063/5.0179854},
    doi = {10.1063/5.0179854},
    number = {2},
    urldate = {2025-05-12},
    journal = {Journal of Applied Physics},
    author = {Li, K. Y. and Li, M. Z. and Wang, W. H.},
    month = jan,
    year = {2024},
    pages = {025102},
}

@article{cassar_designing_2021,
    title = {Designing optical glasses by machine learning coupled with a genetic algorithm},
    volume = {47},
    issn = {02728842},
    url = {http://arxiv.org/abs/2008.09187},
    doi = {10.1016/j.ceramint.2020.12.167},
    number = {8},
    urldate = {2025-05-12},
    journal = {Ceramics International},
    author = {Cassar, Daniel R. and Santos, Gisele G. dos and Zanotto, Edgar D.},
    month = apr,
    year = {2021},
    note = {arXiv:2008.09187 [cond-mat]},
    pages = {10555--10564},
}

@article{debenedetti_supercooled_2001,
    title = {Supercooled liquids and the glass transition},
    volume = {410},
    copyright = {2001 Springer Nature Limited},
    issn = {1476-4687},
    url = {https://www.nature.com/articles/35065704},
    doi = {10.1038/35065704},
    language = {en},
    number = {6825},
    urldate = {2025-05-13},
    journal = {Nature},
    author = {Debenedetti, Pablo G. and Stillinger, Frank H.},
    month = mar,
    year = {2001},
    note = {Publisher: Nature Publishing Group},
    pages = {259--267},
}

@article{binder_molecular_2004,
    title = {Molecular dynamics simulations},
    volume = {16},
    issn = {0953-8984},
    url = {https://dx.doi.org/10.1088/0953-8984/16/5/006},
    doi = {10.1088/0953-8984/16/5/006},
    language = {en},
    number = {5},
    urldate = {2025-05-13},
    journal = {Journal of Physics: Condensed Matter},
    author = {Binder, Kurt and Horbach, Jürgen and Kob, Walter and Paul, Wolfgang and Varnik, Fathollah},
    month = jan,
    year = {2004},
    pages = {S429},
}

@article{yu_structural_2021,
    title = {Structural signatures for thermodynamic stability in vitreous silica: {Insight} from machine learning and molecular dynamics simulations},
    volume = {5},
    shorttitle = {Structural signatures for thermodynamic stability in vitreous silica},
    url = {https://link.aps.org/doi/10.1103/PhysRevMaterials.5.015602},
    doi = {10.1103/PhysRevMaterials.5.015602},
    number = {1},
    urldate = {2025-05-13},
    journal = {Physical Review Materials},
    author = {Yu, Zheng and Liu, Qitong and Szlufarska, Izabela and Wang, Bu},
    month = jan,
    year = {2021},
    note = {Publisher: American Physical Society},
    pages = {015602},
}

@article{yu_stretched_2015,
    title = {Stretched {Exponential} {Relaxation} of {Glasses} at {Low} {Temperature}},
    volume = {115},
    url = {https://link.aps.org/doi/10.1103/PhysRevLett.115.165901},
    doi = {10.1103/PhysRevLett.115.165901},
    number = {16},
    urldate = {2025-05-13},
    journal = {Physical Review Letters},
    author = {Yu, Yingtian and Wang, Mengyi and Zhang, Dawei and Wang, Bu and Sant, Gaurav and Bauchy, Mathieu},
    month = oct,
    year = {2015},
    note = {Publisher: American Physical Society},
    pages = {165901},
}

@article{berthier_modern_2023,
    title = {Modern computational studies of the glass transition},
    volume = {5},
    copyright = {2023 Springer Nature Limited},
    issn = {2522-5820},
    url = {https://www.nature.com/articles/s42254-022-00548-x},
    doi = {10.1038/s42254-022-00548-x},
    language = {en},
    number = {2},
    urldate = {2025-05-13},
    journal = {Nature Reviews Physics},
    author = {Berthier, Ludovic and Reichman, David R.},
    month = feb,
    year = {2023},
    note = {Publisher: Nature Publishing Group},
    pages = {102--116},
}

@misc{mcinnes_umap_2020,
    title = {{UMAP}: {Uniform} {Manifold} {Approximation} and {Projection} for {Dimension} {Reduction}},
    shorttitle = {{UMAP}},
    url = {http://arxiv.org/abs/1802.03426},
    doi = {10.48550/arXiv.1802.03426},
    urldate = {2025-05-15},
    publisher = {arXiv},
    author = {McInnes, Leland and Healy, John and Melville, James},
    month = sep,
    year = {2020},
    note = {arXiv:1802.03426 [stat]},
}

\end{document}